\documentstyle[aps,pre,epsfig]{revtex}
\newcommand{\be}{\begin{equation}}
\newcommand{\ee}{\end{equation}}
\newcommand{\ba}{\begin{eqnarray}}
\newcommand{\ea}{\end{eqnarray}}
\newcommand{\bsigma}{\mbox{\boldmath $\sigma$}}

\newcommand{\br}{{\bf r}}

\newcommand{\bv}{{\bf v}}
\newcommand{\bfc}{{\bf c}}
\newcommand{\hg}{\hat{{\bf g}}}
\newcommand{\bg}{{\bf g}}
\newcommand{\hsigma}{\hat{\bsigma}}
\newcommand{\eps}{\epsilon}

\newcommand{\eno}{^{{\rm E}}}
\newcommand{\enu}{^{{\rm E}}}

\begin{document}
\title{Randomly Driven Granular Fluids: collisional statistics and short
scale structure}
\author{
I.~Pagonabarraga\\
{\it Departament de F\'{\i}sica Fonamental, Universitat de Barcelona, 
Av. Diagonal  647,
08028-Barcelona, Spain}\\
E.~Trizac\\
{\it Laboratoire de Physique Th{\'e}orique\footnote{Unit{\'e}
Mixte de Recherche UMR 8627 du CNRS}, B{\^a}timent 210,
Universit{\'e}  Paris-Sud,
91405 Orsay Cedex, France}\\
T.P.C.~van Noije and M.H.~Ernst\\
{\it Instituut voor Theoretische Fysica, Universiteit Utrecht,
Postbus 80006, 3508 TA Utrecht, The Netherlands}}
\date{\today}
\maketitle
\begin{abstract}
We present a molecular dynamics and kinetic theory study of
granular material, modeled by inelastic hard disks, fluidized by
a random driving force. The focus is on collisional averages and
short distance correlations in the non-equilibrium steady state,
in order to analyze in a quantitative manner the breakdown of
molecular chaos, i.e. factorization of the two-particle
distribution function, $f^{(2)}(x_1,x_2) \simeq \chi f^(1)(x_1)
f^{(1)}(x_2)$ in a product of single particle ones, where  $x_i =
\{{\bf r}_i, {\bf v}_i \}$ with $i=1,2$ and $\chi$ represents the
position correlation. We have found that molecular chaos is only
violated  in a small region of the two-particle phase space $\{
x_1,x_2\}$, where there is a predominance of grazing collisions.
The size of this singular region grows with increasing
inelasticity.  The existence of particle- and noise-induced
recollisions magnifies the departure from mean field behavior.
The implications of this breakdown in several physical quantities
are explored.
\end{abstract}

\section{Introduction}
The interesting phenomena observed in recent experiments with
mono- and
multi-layers of granular material on vibrating plates
\cite{swinney,urbach,gollub,rouyer} show the need to develop
kinetic theories
for rapid granular flows with mechanisms for energy input,
different from those in shear flows or flows through vertical
pipes. In the present article the fluidization is driven by a
random external force, which gives frequent kicks to each particle
in between collisions. Such a driving mechanism has recently been
studied by many authors
\cite{mackintosh,puglisi,cafiero,peng,granmat,pre,moon}.
The basic physical interest is the understanding the
non-equilibrium stationary states (NESS) that exist in the
presence of this random force. The advantage of this fluidization
mechanism, besides its potential physical realizations, lies in
the fact that the NESS is linearly stable against spatial
inhomogeneities.

In Ref.\ \cite{pre}, to which we refer as paper I,
 we have studied the large scale structure and
presented a hydrodynamic description of randomly driven granular
fluids, modeled as systems of smooth inelastic hard spheres (IHS).
The IHS model accounts for two essential features of granular
matter: hard core exclusion and dissipative collisions
\cite{haff}. The dynamics is described by a constant coefficient
$\alpha$ of normal restitution. In collisions a fraction of the
relative kinetic energy is lost, which is proportional to the
inelasticity $\eps=1-\alpha^2$. The stochastic external force
compensates this energy loss, and drives the IHS fluid into a
NESS. This stationary state, though homogeneous and stable
against spatial fluctuations on large space and time scales (at
least for weakly inelastic spheres), was shown to exhibit long
range spatial correlations in density, velocity and granular
temperature fields that extend much beyond the mean free path. In
fact, the corresponding structure functions $S(k)$ diverge as
$1/k^2$ as the wave number $k\to 0$, a behavior caused by the
random external force, which does not conserve momentum whereas
the collisions between particles do. These long range
correlations are of algebraic form, $\sim 1/r^{d-2}$, which
corresponds to $\ln{r}$ in two dimensions ($d=2$). The existence
of such extremely long range spatial correlations is one example
of the many nontrivial properties of non-equilibrium stationary
states in general
\cite{dorfman+kirkpatrick+sengers,grinstein}.

Differences in the stationary states between fluids with
dissipative and conservative interactions also manifest
themselves in the kinetic properties of the fluid, such as the
velocity distribution function, which deviates from a Maxwellian
in particular in the high energy tail of the distribution. In
Ref.\ \cite{granmat} the existence of an overpopulated high
energy tail $\hat{f}\sim \exp[-C v^{3/2}]$, where $C$ is a
constant that depends on the inelasticity, has been obtained from
kinetic theory. A similar behavior has been observed
experimentally at high vibrational accelerations
\cite{gollub,rouyer}. This observation indicates that certain
features of the experiment might be reproduced by modeling the
input of energy into the horizontal motion of the beads by a
random external force, although other energy injection 
mechanisms that could be relevant to
recover the large velocity tail have been put forward \cite{Alain}.
In similar experiments \cite{urbach}
with a vertically vibrating plate covered with a mono-layer of
steel balls with a
packing fraction around 50\%
the velocity distribution of the
horizontal velocities has been measured, and again overpopulated
non-Gaussian high energy tails have been observed.
In the present paper we will investigate the kinetic properties
and short scale correlations
that characterize the stationary state.
More specifically, we will compare
molecular dynamics (MD) simulations of inelastically
colliding disks
with analytic predictions based on the assumption of
molecular chaos.

The Boltzmann equation for dilute gases of particles that
interact via short-ranged repulsive interactions is based on the
assumption of {\it  molecular chaos}, also called the
Stosszahlansatz or mean field approximation. It assumes that the
velocities of colliding particles just before collisions are
uncorrelated, i.e.\ their pair distribution function factorizes,
$\hat{f}^{(2)}(x_1,x_2,t)=\hat{f}(x_1,t)\hat{f}(x_2,t)$, where
$x_i=\{{\bf r}_i,{\bf v}_i\}$ denotes the position and velocity
of particle $i$. Enskog's extension of the Boltzmann equation to
a dense system of hard spheres \cite{chapman}, referred to as
Enskog-Boltzmann equation, is also based on the fundamental
assumption of the absence of velocity correlations. Here the
assumption of molecular chaos postulates that
$\hat{f}^{(2)}(x_1,x_2,t)=\chi \hat{f}(x_1,t)\hat{f}(x_2,t)$ for
approaching particles (${\bf v}_{12}\cdot {\bf r}_{12}<0$) {\it
just before} collision ($r_{12} =\sigma+0$), where $\chi$ is
assumed to be the radial pair distribution function at contact
$g(r_{12}=\sigma+0)$ in local equilibrium. It implies the {\it
additional} assumption that spatial correlations between
colliding particles just before collision are independent of
their velocities, i.e. absence of position-velocity correlations.
The Enskog $\chi$ factor enhances the collision frequency at
higher densities. For dilute gases the assumption of molecular
chaos seems to be justified. Recently Lutsko \cite{Lutsko} and
Soto and Mareschal \cite{SM} derived for a freely evolving
inelastic hard disk fluid a relation between pre- and
 post-collision radial distribution function at contact, as a
 function of the angle, $\theta = \cos^{-1} (\hat{{\bf v}}_{12}
 \cdot \hat{{\bf r}}_{12} )$,
between the relative velocity ${\bf v}_{12}$ of the colliding
 particles, and their relative position at contact ${\bf r}_{12}$,
and they confirmed their results by MD simulations. Their
observations made it clear that further arguments are needed to
clarify the meaning of the $\chi-$factor in Enskog's formulation
of the molecular chaos assumption. This will be done in section
II A.

The breakdown of molecular chaos at higher densities in classical
fluids with conservative forces has been extensively investigated
in the sixties and seventies \cite{60-70}. This breakdown is
caused by sequences of correlated binary collisions, the so
called ring collisions \cite{berne}. They lead to long time tails
in velocity and stress autocorrelation functions
\cite{dorfman+cohen,ernst+hauge+vleeuwen}, and to long range
spatial correlations in NESS \cite{dorfman+kirkpatrick+sengers}.
The quantitative effects of velocity correlations on transport
coefficients at liquid densities are also significant. For
instance, molecular dynamics simulations on elastic hard sphere
systems at liquid densities \cite{hansen} have shown that the long
time tails increase the measured self-diffusion coefficient $D$
typically by 15\% to 20\% with respect to the prediction of the
Enskog theory, $D_{\rm E}=D_{\rm B}/\chi$, where $D_{\rm B}$ is
the Boltzmann value of the self-diffusion coefficient.

A well-known example of short scale structure
in granular fluids are the position-velocity correlations leading to
the phenomenon of inelastic
collapse \cite{mcnamara,EPJE}, which is  a
{\it divergence} of the collision frequency $\omega$
in a finite time. The
collapse singularity implies that an infinite number of
collisions occurs within a finite time interval in a subset of
(nearly) touching particles, ordered in linear arrays.
The phenomenon is, however, an artifact of the assumption that the
coefficient of restitution $\alpha$ is independent of the impact
velocities, whereas on physical grounds $\alpha(v_{12})\to 1$
(elastic limit), as the relative velocity $v_{12}$ vanishes.
Molecular dynamics simulations have shown that the
assumption of molecular chaos is also violated
in undriven granular fluids in their late stages of evolution,
the so called nonlinear clustering regime.
For instance, the measured distribution of impact parameters is not
uniform, as expected on the basis of molecular chaos, but biased
toward grazing collisions \cite{muller,sluding,Gold}.

As shown below, in the driven IHS fluid there is an important
additional
reason for the breakdown of molecular chaos, namely the strong
increase in  collision frequency at small relative velocities
between two isolated particles, caused by the so-called noise
induced re-collisions. This
correction to the collision frequency, that is important at all
densities, is  also neglected in the molecular chaos assumption.

The main goal of this paper is to quantify, analyze and interpret
the effects of the breakdown of molecular chaos in the NESS of
inelastic hard spheres that are subject to a random external
force between collisions. We will focus in particular on
velocity-velocity correlations and position-velocity correlations
between particles almost in contact, i.e.\ the short scale
structure.

Section \ref{sec:predictions} presents the analytic results,
based on the Enskog-Boltzmann equation, which has been modified
to account for the external energy input. In section \ref{sec:md}
we present molecular dynamics results for several quantities that
characterize the collision processes and related short scale
structure of the NESS, and make a comparison with predictions
based on molecular chaos.

%%%%%%%%%%%%%%%%%%%%%%%%%%%%%%%%%%%%%%%%%%%%%%%%%%%%%%%%%%%%%%
\section{Kinetic theory for the NESS}
\label{sec:predictions}
\subsection{Molecular chaos and Enskog approximation}
\label{subsec:mc}
The Enskog-Boltzmann equation for the single-particle
distribution $\hat{f}(\bv_1,t)$ in a spatially homogeneous
randomly driven fluid of inelastic hard spheres of diameter
$\sigma$ reads in $d=2$ or 3 dimensions \cite{granmat}:
\ba
&\partial_t \hat{f}(\bv_1,t)=n \chi \sigma^{d-1} \int {\rm d}{\bf
v}_2 \int {\rm d}\hat{\bsigma} \Theta ({\bf
v}_{12}\cdot\hat{\bsigma})
({\bf v}_{12}\cdot\hat{\bsigma}) \times & \nonumber\\
& \left\{\frac{1}{\alpha^2}\hat{f}({\bf
v}_1^{\ast\ast},t)\hat{f}({\bf v}_2^{\ast\ast},t)
 -\hat{f}({\bf v}_1,t)\hat{f}({\bf v}_2,t) \right\} +
 \frac{\xi_0^2}{2} \left(\frac{\partial}{\partial{\bf v}_1}\right)^2
 \hat{f}(\bv_1,t)&,
\label{eq:fp}
\ea
where ${\bf v}_{12}={\bf v}_1-{\bf v}_2$ and $n$ the number
density. The Heaviside function $\Theta (x)$ restricts the
$\hat{\bsigma}$ integration to the hemisphere ${\bf
v}_{12}\cdot\hat{\bsigma}>0$, where $\hat{\bsigma}$ is the unit
vector along the line of centers of the colliding spheres at
contact, pointing from particle 2 to 1. In the sequel $\hat{{\bf
a}} = {\bf a}/|{\bf a}|$ denotes a unit vector. The gain term of
the collision integral describes the {\em restituting} collisions
that convert the precollision velocities
($\bv_1^{\ast\ast},\bv_2^{\ast\ast}$) into ($\bv_1,\bv_2$), while
the loss term describes the {\em direct} collisions, and contains
the precollision velocities $( {\bf v}_1,{\bf v}_2)$ leading to
postcollision velocities $( {\bf v}^*_1,{\bf v}^*_2)$. The
postcollision and restituting velocities have been defined in
\cite{vleeuwen}. The $\chi$ factor will be discussed below.

As derived in \cite{granmat}, the Fokker-Planck term accounts for
the external energy input, and describes
diffusion in velocity space with a diffusivity
proportional to the rate of energy input $ \xi_0^2$ per unit mass.
Here $\xi_0$ is the strength of the random driving force, which
is assumed to be Gaussian white noise \cite{granmat,pre}.

Before studying the short scale structure, we consider the single
particle distribution function $\hat{f}(\bv)$ in the NESS. The
stationary solution of    (\ref{eq:fp}) is determined by the
balance, $ m\xi_0^2 = \Gamma$, of external heating, $m \xi_0^2$,
and internal loss of energy due to collisions, $\Gamma$. It is
characterized by a time independent temperature, $T = \langle
 m v^2  /d\rangle$,  defined as the average kinetic energy per
particle, and discussed in paper I. As mentioned in the
introduction, this stationary solution exhibits an overpopulated
high energy tail $\hat{f}\sim
\exp[-C v^{3/2}]$.
The structure of the tail distribution is determined by collisions of
very energetic particles with `thermal' particles, and can be
obtained by neglecting the gain term in the Boltzmann equation
\cite{granmat}.

In Ref.\ \cite{granmat} $\hat{f}(\bv)$ has been calculated by
solving the Enskog-Boltzmann equation    (\ref{eq:fp}) by an
expansion in Sonine polynomials. To formulate this result it is
convenient to introduce a rescaled distribution function
${f}({\bf c})$, defined by $\hat{f}({\bf v})\equiv[1/v_0^d] f({\bf
c})$ with ${\bf c}\equiv \bv/v_0$, where $v_0\equiv\sqrt{2T/m}$
is the thermal velocity and $d$ the dimensionality. This gives
\be
f({\bf c})=\varphi(c)\left\{1+a_2
\left[\frac{1}{2}c^4-\frac{1}{2}(d+2)c^2+\frac{1}{8}d(d+2)\right]
+\dots\right\},
\label{eq:expansion}
\ee
where  the Maxwellian $\varphi(c)\equiv \pi^{-d/2}\exp(-c^2)$.
Note that $a_2$ is proportional to the fourth cumulant
of the scaling form $f({\bf c})$, i.e.\
\be
a_2=\frac{4}{d(d+2)}\left[\langle
c^4\rangle-\textstyle{\frac{1}{4}}d(d+2)\right]=
\textstyle{\frac{4}{3}}\left[\langle c_x^4\rangle -3{\langle
c_x^2\rangle}^2\right]
\label{eq:cumul},
\ee
and vanishes in the elastic limit.
An explicit calculation to linear order in $a_2$ gives
\cite{granmat}
\be
a_2=\frac{16(1-\alpha)(1-2 \alpha^2)}{73
+56 d-24\alpha d -105 \alpha + 30(1-\alpha) \alpha^2}.
\label{eq:a2}
\ee
In the next section this prediction will be tested against
molecular dynamics simulations.

 Consider first the exact expression for the mean collision frequency in
 the homogeneous NESS, defined as
\be
\omega=n \sigma^{d-1} \int {\rm d}{\bf v}_1 \int{\rm d}{\bf v}_2
\int{\rm d}\hat{\bsigma} \Theta (-{\bf v}_{12}\cdot\hat{\bsigma})
|{\bf v}_{12}
\cdot \hat{\bsigma}|  \hat{f}^{(2)}({\bf v}_1,{\bf v}_2,\bsigma),
\label{eq:omegadef}
\ee
where $ \hat{f}^{(2)}({\bf v}_1,{\bf v}_2,{\bf \sigma})$
 is the {\it dynamic} or {\it constrained }pair distribution
 function with velocities aiming to collide, just {\it before} contact
 with $ {\bf r}_{12}=\bsigma$. {\em Molecular chaos},
 also referred to as mean  field theory, requires the complete
 factorization of the dynamic precollisional pair function,
\be
 \hat{f}^{(2)}({\bf v}_1,{\bf v}_2,\bsigma)=\chi
 \hat{f}({\bf v}_1)  \hat{f}({\bf v}_2).
\label{eq:mc}
\ee
What is the meaning of the $\chi-$factor, used in formulating the
{\it molecular chaos} hypothesis? This hypothesis for {\em dilute
gases} concerns the absence of correlations in {\it precollision}
velocities, and  in {\it precollision} positions $(\chi = 1)$. In
{\em dense fluids} on the other hand,
  the precollision position correlation $\chi$ is different from 1,
 but the precollision velocity-velocity and position-velocity
correlations are still assumed to be absent.

In the literature it is common to take $\chi$  equal to the
radial distribution function at contact in {\it local
equilibrium}, i.e. $\chi = \chi_{\rm E} \equiv g_{\rm eq}(r \to
\sigma+0)$, which mainly accounts for precollision hard core
exclusion effects. For hard disks and hard spheres the latter
function is approximately given by the Verlet-Levesque (2D) and
Carnahan-Starling (3D) approximations
\cite{verlet},
\ba
\chi_{\rm E}(\phi)&=&(1-\textstyle{\frac{7}{16}}\phi)
/(1-\phi)^2\,\,\,\,\,\mbox{(2D)}
\nonumber\\
\chi_{\rm E}(\phi)&=&(2-\phi)/2(1-\phi)^3\,\,\,\,\,\mbox{(3D)},
\label{eq:vlcs}
\ea
where the packing fraction  in $d$ dimensions is defined as
$\phi=n(\sigma/2)^d \Omega_d/d$, and $\Omega_d=2
\pi^{d/2}/\Gamma(d/2)$ is the surface area of a $d$-dimensional
unit sphere. In this paper we refer to the molecular chaos
approximation with   $\chi = \chi_{\rm E}$, as the {\it Enskog
approximation}.

In principle, different options are open for the $\chi-$factor.
As $\hat{f}^{(2)}$ is the dynamic precollision pair distribution
function at contact, an alternative choice for the $\chi$ in the
factorized form could be the dynamic precollision  radial
distribution function at contact, defined as an average over the
precollision hemisphere,
\be \label{eq:g-min}
\chi^{(-)} \equiv [2/\Omega_d] \int d {\bf v}_1 \int  d{\bf v}_2
\int{\rm d}\hat{\bsigma} \Theta (-{\bf v}_{12}\cdot\hat{\bsigma})
\hat{f}^{(2)}({\bf v}_1,{\bf v}_2,\bsigma).
\ee
Another option could be the unconstrained radial distribution, $
g(r)$, in the NESS, extrapolated to contact $(r \to \sigma)$.
This function is further discussed in subsection II D.

For the randomly heated fluid under study here, the dynamics is
not purely hard-sphere like. The random force acting on the
particles may be expected  to contribute to the value of the pair
distribution function at contact. This effect will be addressed
in the subsequent sections.

Equation (\ref{eq:omegadef}) with $ \hat{f}({\bf v})$ replaced by the
Maxwellian,  yields for the collision
 frequency in the molecular chaos approximation
 $ \omega_{\rm mc}(T) = \chi \omega_0 (T) $ , and more
 specifically in the Enskog approximation,
\be
\omega_{\rm E}(T) = \chi_{ \rm E}\omega_0 (T).
\label{eq:omega-Ensk}
\ee
Here the Boltzmann collision frequency for dilute gases is given
by,
\be
\omega_0 (T) = \Omega_d  n \sigma^{d-1} \sqrt{ T/\pi m},
\label{eq:omega-boltz}
\ee
and the small correction of ${\cal O}(a_2)$ appearing in
   (\ref{eq:expansion}) has been neglected.

Spatial correlation functions in non-equilibrium stationary states
are quite different from
local equilibrium ones, and show long range correlations due to
correlated sequences of ring collisions, also referred to as mode
coupling effects \cite{dorfman+kirkpatrick+sengers}. In paper I
we have shown the existence of very {\em long}
range correlations $\sim 1/r^{d-2}$ in the randomly driven IHS fluid.
The {\em short} range correlations in the NESS can in principle be
obtained from the ring kinetic equation for IHS (see Ref.\
\cite{vleeuwen}).
However, systematic methods to evaluate collision integrals and
pair correlation functions at short distances using this ring
kinetic theory have not yet been developed. In the section on
simulation results, we return to the effects of ring collisions,
and present  arguments why their contributions are expected to be
more important here than for elastic hard spheres.

%%%%%%%%%%%%%%%%%%%%%%%%%%%%%%%%%%%%%%%%%%%%%%%%%%%%%%%%%%%%%%%%%%%%%%%%%
%
\subsection{Collisional averages}
In hard sphere systems there are many properties that involve the
pair distribution function of particles just before collision. To
study these, it is convenient to introduce the {\em collisional}
average $\langle ...\rangle_{\hbox{\scriptsize coll}}$ for a
quantity $A$ in the NESS, defined as
 \be
\langle A({\bf v}_1,{\bf v}_2,\bsigma)\rangle_{\hbox{\scriptsize coll}} =
\frac{ \int {\rm d}{\bf v}_1 \int{\rm d}{\bf v}_2
\int{\rm d}\hat{\bsigma} \Theta (-{\bf v}_{12}\cdot\hat{\bsigma})|
{\bf v_{12}} \cdot \hat{\bsigma}| A({\bf v}_1,{\bf v}_2,\bsigma)
\hat{f}^{(2)}({\bf v}_1,{\bf v}_2,\bsigma)}
{\int {\rm d}{\bf v}_1 \int{\rm d}{\bf v}_2 \int{\rm d}
\hat{\bsigma} \Theta (-{\bf v}_{12}\cdot\hat{\bsigma}) |{\bf v}_{12}\cdot
\hat{\bsigma}| \hat{f}^{(2)}({\bf v}_1,{\bf v}_2,\bsigma)}.
\label{eq:collav}
\ee
In the sequel it is more convenient to work with a rescaled pair
distribution function $\hat{f}^{(2)}({\bf v}_1,{\bf
v}_2,\br_{12})=[1/v_0^{2d}]$ $ f^{(2)}({\bf c}_1,{\bf
c}_2,\br_{12})$. To express the collisional averages
(\ref{eq:collav}) in rescaled variables one replaces
$\hat{f}^{(2)}$ by $f^{(2)}$, ${\bf v}_i$ by ${\bf c}_i$, ${\bf
v}_{12}$ by ${\bf g}={\bf c}_1-{\bf c}_2$, and
 $A({\bf v}_1,{\bf v}_2,\bsigma)$ by
 $A(v_0 {\bf c}_1,v_0 {\bf c}_2,\bsigma)$.

These objects can be conveniently computed in event driven
molecular dynamics algorithms for hard sphere systems
\cite{linked}. Collisional averages are defined for particles
that are about to collide (i.e. $|r_{12}|=\sigma+0$), and can be
calculated from kinetic theory using the molecular chaos
assumption, possibly supplemented with the Enskog approximation
at higher densities.

Collisional averages of great importance are the collisional
energy loss per unit time, $\textstyle{ \frac{d}{2}} n \Gamma$,
and the excess hydrostatic pressure, $p-n T$, resulting from
collisional transfer of momentum. With a minor generalization to
$d$ dimensions we obtain from Ref.\cite{brey+dufty+santos} the
exact expression for the pressure in the NESS:
\ba
\frac{p(T)}{n T}-1&=&\left(\frac{1+\alpha}{2d}\right) n\sigma^d
\int {\rm d}{\bf c}_1 \int{\rm d}{\bf c}_2
\int{\rm
d}\hat{\bsigma} \Theta (-{\bf g}\cdot\hat{\bsigma}) |{\bf
g}\cdot
\hat{\bsigma}|^2  f^{(2)}({\bf c}_1,{\bf c}_2,\bsigma)
\nonumber\\
&=&\left(\frac{1+\alpha}{2d}\right) \frac{\sigma \omega}{v_0} \langle
|{\bf g}\cdot \hat{\bsigma}|\rangle_{\hbox{\scriptsize coll}}  .
\label{eq:press}
\ea
The second equality is obtained by introducing the collisional
average (\ref{eq:collav}) and expressing its denominator in terms
of the collision frequency given by     (\ref{eq:omegadef}). In fact,
inserting (\ref{eq:mc}) into the first line of (\ref{eq:press})
allows one to carry out the $\hat{
\sigma}-$integration, and the right hand side becomes
proportional to the rescaled velocity average $ \int {\rm d}{\bf
c}_1 \int{\rm d}{\bf c}_2 g^2 f( {\bf c}_1) f({\bf c}_2) =2$
without any further assumption about neglecting the term
proportional to $a_2$ in (\ref{eq:expansion}). This argument is
special for the pressure, as other moments involve the knowledge
of the complete $f(c)$. Indeed, the generic collisional average 
becomes $\langle |{\bf g}\cdot \hat{\bsigma}|^m
\rangle_{\hbox{\scriptsize coll}}  = 2^{m/2} \;
\Gamma ( \frac{1}{2} m +1)$, independent of
dimensionality, assuming molecular chaos (\ref{eq:mc}) and replacing $f(c)$ 
by the Maxwellian $\varphi(c)$ (the contributions coming from $a_2$ are
quite small and can be neglected; they have been computed in \cite{granmat}).
Finally, the pressure can be expressed as,
 \be
\frac{p_{\rm mc}(T)}{nT}-1=2^{d-2}(1+\alpha) \chi \phi.
\label{eq:pressmc}
\ee
Different choices for $\chi$ yield different approximations. For
instance, with $\chi = \chi_{\rm E}$ we obtain the Enskog
approximation $p_{\rm E}(T)$ for the pressure of IHS.

In the elastic limit $p_{\rm E}(T)$ at $\alpha=1$ gives the
standard equation of state for elastic hard disks or spheres.
Notice that the pressure for IHS is only defined {\it
kinetically} as the momentum flux, which leads to
(\ref{eq:press}).  A statistical mechanical derivation of the
equation of state from the partition function or free energy  for
the IHS fluid does not exist.

In a similar manner we obtain the exact expression for the
collisional damping rate,
\ba
\Gamma (T)&=&\left(\frac{1-\alpha^2}{2d}\right)n
\sigma^{d-1}v_0 T
\int {\rm d}{\bf c}_1 \int{\rm d}{\bf c}_2
\int{\rm
d}\hat{\bsigma} \Theta (-{\bf g}\cdot\hat{\bsigma}) |{\bf
g}\cdot
\hat{\bsigma}|^3  f^{(2)}({\bf c}_1,{\bf c}_2,\bsigma)
\nonumber\\
&=& \gamma_0 \omega T \langle |{\bf g}\cdot
\hat{\bsigma}|^2\rangle_{\hbox{\scriptsize coll}} \nonumber\\
&=& m \,\xi_0^2,
\label{eq:gammadef}
\ea
where $\gamma_0=(1-\alpha^2)/2d$ is the dimensionless damping
constant introduced in Refs.\ \cite{granmat,pre}. The last
equality (\ref{eq:gammadef}) expresses the balance between the
energy input due to the white noise, and the collisional loss of
energy in the NESS, and determines the temperature $T$ in the
NESS. By specializing this equation to the Enskog mean field
approximation, $f^{(2)} = \chi_{\rm E} f f$, where $\langle |{\bf
g}\cdot \hat{\bsigma}|^2\rangle_{\hbox{\scriptsize coll}}=2$, we
obtain the approximate result,
 \be
\Gamma_{\rm E}(T)=2 \gamma_0 \omega_{\rm E} (T) T,
\label{eq:gamma-Ensk}
\ee
and similar relations for different choices of $\chi$. It is
convenient to define a {\it reference} temperature $T_{\rm E}$
through the relation,
\be
\Gamma_{\rm E}(T_{\rm E}) = m\xi_0^2,
\label{T-Ensk-implicit}
\ee
or more explicitly
\be
T_{\rm E} = m \left(\frac{\xi_0^2 \sqrt{\pi}}{2 \gamma_0
\Omega_d \chi_{\rm E} n
\sigma^{d-1}}\right)^{2/3}.
\label{eq:T-Ensk}
\ee
 Moreover the definition of $T_{\rm E}$ combined with the NESS
condition $\Gamma(T) =m \xi_0^2$  implies the relation $\Gamma(T)
=\Gamma_{\rm E} (T_{\rm E})$, and consequently,
 \be
 \frac{\Gamma(T)}{\Gamma_{\rm E}(T)} =
 \frac{\Gamma_{\rm E}(T_{\rm E})}{\Gamma_{\rm E}(T)}
  = \left(\frac{T_{\rm E}}{T}\right)^{3/2}.
  \label{eq:ratio-gamma}
\ee
 In the sequel
  we will also use a reference frequency $\omega_{\rm E}$
  without any argument to denote,
\be
\omega_{\rm E} \equiv \omega_{\rm E}(T_{\rm E}) =
\chi_{\rm E} \Omega_d  n \sigma^{d-1}
\sqrt{T_{\rm E}/\pi m}.
\label{omega-Ensk}
\ee
Although we will discuss the simulation results in detail in the
next section, it is of interest already at this point to note
that for these systems, as shown in Fig.
\ref{fig:temp_omega}, the ratio of the kinetic temperature and
the reference temperature, $T/T_{\rm E}$, is only somewhat larger
than 1 for all $\alpha$, that it approaches 1 in the elastic
limit $(\alpha
\to 1)$, and that it monotonically increases with decreasing
$\alpha$ (see Fig. \ref{fig:temp_omega}). The same figure shows that the ratio,
$\omega/\omega_{\rm E}$ also approaches 1 for $\alpha
\to 1$, with a steep increase to a value 5.6 as $\alpha \to 0$.
Further discussion of these points is postponed till section III.
If the Enskog factorization, $\hat{f}^{(2)}= \chi_{\rm E}
\hat{f}
\hat {f} $, would be exact, then $T=T_{{\rm E}}$ and
$\omega=\omega_{{\rm E}}$.

A third quantity of interest, the precollisional $\chi^{(-)}$
factor, defined in (\ref{eq:g-min}),  can also be expressed as a
collisional average using (\ref{eq:collav}),
\be \label{eq:g-min-coll}
\chi^{(-)} = ( 2 \omega / \Omega_d n \sigma^{d-1} v_0)
\langle|{\bf g} \cdot \hsigma|^{-1} \rangle_{\hbox{\scriptsize
coll}}.
\ee

Before closing this section a {\it caveat} about internal
consistency is appropriate. To obtain consistent theoretical
predictions for the pressure $p$ or dissipation rate $\Gamma$, it
is paramount that both factors, $\omega$ and $ \langle | {\bf g}
\cdot {\bf \sigma}|^m \rangle_{\hbox{\scriptsize coll}}   $, be
calculated using identical approximations for $  f^{(2)} $. For
instance,
 the mean field or molecular chaos approximation for the
 dissipation rate, $\Gamma_{{\rm E}}(T) = 2 \gamma_0 \omega T$,
-- an expression  commonly used in granular hydrodynamic
equations -- should
necessarily  be combined with $\omega_{{\rm E}}(T)$ in
    (\ref{eq:omega-Ensk}). Any improved
theoretical calculation for $\omega$ without a concomitant
correction to the mean field result for $ \langle | {\bf g} \cdot
{\bf \sigma}|^m \rangle_{\hbox{\scriptsize coll}} $ is {\it
inconsistent  }.

%%%%%%%%%%%%%%%%%%%%%%%%%%%%%%%%%%%%%%%%%%%%%%%%%%%%%%%%%%%
\subsection{Velocity distributions}
\label{subsec:vel}
We study a variety of collisional averages  and corresponding
probability distributions. By choosing $A({\bf c}_1,{\bf
c}_2,\bsigma)=\delta(|{\bf g}|-g)$ we obtain the probability
$P_r(g)$ that two colliding particles have a relative speed
$|{\bf c}_{12}|=g$.  From here on we only quote results for two
dimensions. {\em Analytic} calculations are based on the
molecular chaos assumption (\ref{eq:mc}) in combination with
(\ref{eq:expansion}). Inspection of (\ref{eq:collav}) shows that
under this assumption the collisional averages are independent of
the $\chi-$factor. Straightforward algebra gives for the
constrained $g-$distribution,
\ba \label{b1}
P_r(g)&=& \langle \delta(|{\bf c}_{12}|-g)
\rangle_{\hbox{\scriptsize coll}}
\nonumber\\
&=&\sqrt{\frac{2}{\pi}}
g^2 e^{-\textstyle{\frac{1}{2}}g^2}
\left\{1+\frac{1}{16}a_2(
g^4-8 g^2+9)\right\}.
\ea
Similarly we obtain the probability distribution for the center
of mass velocity, ${\bf G}\equiv
\textstyle{\frac{1}{2}}({\bf c}_1+{\bf c}_2)$,
\label{eq:b2}
 \ba
P_{{\rm CM}}(G)&=& \langle \delta(|{\bf C}|-G)
\rangle_{\hbox{\scriptsize coll}}
\nonumber\\
&=& 4 G e^{-2 G^2}\left\{1+a_2(G^4-G^2)\right\}.
\ea
It equals the unconstrained  equilibrium distribution function
apart from a small term of ${\cal O}(a_2)$.
Furthermore, the probability that the precollision  speed
$|{\bf c}_i|$ of one of the
colliding particles $(i=1,2)$ has a value $v$ is
\ba
P(v)&=& \langle \delta(|{\bf c}_1|-v)
\rangle_{\hbox{\scriptsize coll}}  \nonumber\\
&=& \sqrt{2} v e^{-3 v^2/2} \left\{(1+v^2)
I_0\left(\textstyle{\frac{1}{2}}v^2\right)
+v^2I_1\left(\textstyle{
\frac{1}{2}} v^2\right)\right\},
\label{eq:Pc_1}
\ea
whereas the unconstrained distribution is $\sim v \exp(-v^2)$.
In evaluating this collisional average we neglect the $a_2$
contribution, and carry out the constrained $\hat{\bsigma}$
integration.
To calculate the remaining integral $\int{\rm d}{\bf c}_2 c_{12}
\varphi(c_2)$, we change integration variables to ${\bf
g}$ expressed in polar coordinates $\{g,\phi\}$, and use
the relation $\int_0^\pi {\rm d}\phi
\exp(-2 c_1 g \cos{\phi})=\pi I_0(2 c_1 g)$.
The subsequent $g-$integration follows from    (6.618.4) in
Ref.\cite{gradshteyn}. Using the asymptotic expressions
$I_0(x)\sim I_1(x)\sim
\exp(x)/\sqrt{2\pi x}$ for the modified Bessel functions of
the zeroth and first order, we obtain the high energy behavior,
$P(v)\sim 2\sqrt{2/\pi} v^2 \exp(-v^2)$.

In a similar manner we obtain the following velocity moments and
correlations, using the molecular chaos assumption,
\ba
\langle g^2\rangle_{\hbox{\scriptsize coll}}
&=&3\{1+\textstyle{\frac{1}{4}}a_2\},\,\,\,\,\,
\langle G^2\rangle_{\hbox{\scriptsize coll}}  =
\textstyle{\frac{1}{2}}\{1+\textstyle{\frac{1}{2}}a_2\},
\nonumber\\
\langle g^{\ast 2}\rangle_{\hbox{\scriptsize coll}}  &=&\langle
g^2\rangle_{\hbox{\scriptsize coll}}
-(1-\alpha^2)\{1+\textstyle{\frac{1}{4}}a_2\},\nonumber\\
\langle c_1^2\rangle_{\hbox{\scriptsize coll}}   &=&   \langle
G^2\rangle_{\hbox{\scriptsize coll}}   +
\textstyle{\frac{1}{4}}\langle g^2
\rangle_{\hbox{\scriptsize coll}}   =
\textstyle{\frac{5}{4}}\{1+
\textstyle{\frac{7}{20}}a_2\},\nonumber\\
\langle {\bf c}_1\cdot{\bf c}_2
\rangle_{\hbox{\scriptsize coll}}  &=&
\langle G^2\rangle_{\hbox{\scriptsize coll}}   -
\textstyle{\frac{1}{4}}\langle g^2
\rangle_{\hbox{\scriptsize coll}}
=-\textstyle{\frac{1}{4}}\{1-\textstyle{\frac{1}{4}}a_2\},
\nonumber\\
\langle {\bf c}_1^\ast\cdot{\bfc}_2^\ast
\rangle_{\hbox{\scriptsize coll}}
&=&\langle {\bf c}_1\cdot{\bf c}_2
\rangle_{\hbox{\scriptsize coll}}  +
\textstyle{\frac{1}{2}}(1-\alpha^2)
\{1+\textstyle{\frac{1}{4}}a_2\}.
\label{eq:averages}
\ea
Here ${\bf c}_i^\ast$ are the postcollision velocities, as
defined in paper I.  The sum of the third and fourth equality
depends only on the center of mass velocity, i.e. $\langle G^2
\rangle_{\hbox{\scriptsize coll}}  $.
%The {\em elastic limit}, where $\alpha\to 1$ or
%$\eps=1-\alpha^2 \to 0$, refers to the procedure of
%simultaneously reducing
%the energy input $\sim \xi_0^2$ and the inelasticity $\eps$ at a
%constant ratio, such that the system remains in a NESS,
%that approaches the equilibrium state of elastic hard spheres.
In the elastic limit $\alpha\to 1$, the average
energy of a particle
that is about to collide, $\langle c^2
\rangle_{\hbox{\scriptsize coll}}  =(5/4)
\langle c^2 \rangle$,
is above the mean energy per particle, $\langle c^2 \rangle$,
which equals unity.

In the molecular chaos approximation an average like
   $\langle ({\bf c}_1\cdot{\bf c}_2)^m g^n
\rangle_{\hbox{\scriptsize coll}}$ with  $\{m,n\}$
integers, is in general non-vanishing, except in the special case
$n=-1$. Then $\langle ({\bf c}_1\cdot{\bf c}_2)^m /g
\rangle_{\hbox{\scriptsize coll}}  $
reduces to an unconstrained average, proportional $\langle ({\bf
c}_1\cdot{\bf c}_2)^m \rangle $, which vanishes for odd values of
$m$. Additional information about the relative orientation of the
incoming velocities can be obtained from the distribution of the
angle $\psi_{12}$, defined by ${\bf c}_1\cdot {\bf c}_2=c_1 c_2
\cos{\psi_{12}}$. A numerical calculation (again neglecting $a_2$
corrections) gives $\langle
\cos{\psi_{12}}\rangle_{\hbox{\scriptsize coll}}
\simeq -0.233$,
which is close to the value $-0.2$, estimated
from $\langle {\bf c}_1\cdot{\bf c}_2
\rangle_{\hbox{\scriptsize coll}}   \simeq
\langle
c_1^2\rangle_{\hbox{\scriptsize coll}}   \langle \cos{\psi_{12}}
\rangle_{\hbox{\scriptsize coll}}  $ using the above
results.

A very sensitive probe for studying the violation of molecular
chaos is the probability distribution $P(b)$ of impact parameters,
$b = |\hg \times \hsigma| = \sin\theta$, where $ \theta =
\cos^{-1}(\hg \!\cdot\! \hsigma)$ is the angle of incidence.
It is defined as the collisional average
\be \label{a12}
P(b) = \langle \delta \left(b - |\hg \times \hsigma|\right)
\rangle_{\hbox{\scriptsize coll}}
=\frac{\int d \hsigma \int d{\bf c}_1 \int d{\bf c}_2 \,
\delta(b-|\hg \times \hsigma|)\, |{\bf g} \cdot \hsigma|\,
 \Theta (- \hg \cdot \hsigma ) f^{(2)}({\bf c}_1,{\bf c}_2,\bsigma)}
 { \int d \hsigma \int d{\bf c}_1
 \int d{\bf c}_2\, |{\bf g} \cdot \hsigma| \,
 \Theta (- \hg \cdot \hsigma )
 f^{(2)}({\bf c}_1,{\bf c}_2,\bsigma )},
\ee
and $P(b)$ can be easily computed in a molecular dynamics
experiment. As long as molecular chaos holds, the distribution of
$b$ is independent of the functional form of $ f$ and we obtain
straightforwardly
\be
P(b)=\left\{
\begin{array}{cl}
(d-1)\, b^{d-2} & \mbox{if $0<b<1$}\\
0 & \mbox{otherwise},
\end{array}\right.
\label{b33}
\ee
which reduces in two dimensions to the uniform distribution,
\be
P(b)=\left\{
\begin{array}{ll}
1&\mbox{if $0<b<1$}\\
0&\mbox{otherwise}.
\end{array}\right.
\label{b3}
\ee

In order to analyze molecular chaos breakdown in more detail,
we have introduced a collection of moments $M_{nm}$ and their
dimensionless counterparts $B_{nm}$ for $n,m=\{0,1,2 \dots \}$,
(see definition below),  to analyze in detail the possible
breakdown of the molecular chaos factorization (\ref{eq:mc}).
These moments $M_{nm}(T)$ of the pair distribution at contact are
defined as,
\be  \label{Mnm}
M_{nm} (T) \equiv \frac{2}{\Omega_d} \int d {\bf v}_1
d{\bf v}_2 \int d\hat{\bsigma} \Theta (- {\bf v}_{12} \cdot {\hsigma} )
 \hat{f}^{(2)}({\bf v}_1,{\bf v}_2,\bsigma|T)\,\,
 v_{12}^n \,|\cos \theta|^m ,
 \ee
which are  averages over the precollision hemisphere, where $\theta =
\cos^{-1} (\hg \cdot \hsigma)$. Let $M^{\rm
E} _{nm} (T_{\rm E})$ denote  the same quantity evaluated in
Enskog's formulation of the {\it molecular chaos} approximation,
{\it and} evaluated at the reference temperature $T_{\rm E}$, i.e.
evaluated with $\hat{f}^{(2)}({\bf v}_1,{\bf v}_2,\bsigma|T)$
replaced by $\chi_{\rm E} \hat{f}({\bf v}_1|T_{\rm E})\hat{f}({\bf
v}_2|T_{\rm E})$, then the reduced moments are  defined as
\be \label{a4}
B_{nm}(T) =M_{nm}(T)/ M^{{\rm E}}_{nm}(T_{\rm E}),
\ee
where $M^{{\rm E}}_{nm}(T_{\rm E})$ is  evaluated in
(\ref{MEnm}). It is proportional to $v_{\rm E}^n$, where $v_{\rm
E} = \sqrt{2T_{\rm E}/m}$. We prefer to normalize the reduced
moments by  $M^{{\rm E}}_{nm}(T_{\rm E})$, because its analytic
form is given explicitly. One could also normalize by $M^{{\rm
E}}_{nm}(T)= (T_{\rm E}/T)^{n/2}M^{{\rm E}}_{nm}(T_{\rm E}) $. The
disadvantage of $M^{{\rm E}}_{nm}(T)$ is that the computation
requires the simulated values of the kinetic temperature $T$. The
collisional averages $\langle v_{12}^n |\cos\theta|^m
\rangle_{\hbox{\scriptsize coll}}$ expressed in terms of
these new moments give,
 \be \label{a1}
\langle v_{12}^n |\cos\theta|^m\rangle_{\hbox{\scriptsize coll}}
= \frac{M_{n+1,m+1}(T)}{ M_{11}(T)}.
\ee
We first observe that the average collision frequency $\omega$,
defined in (\ref{eq:omegadef}), is  proportional to $M_{11} (T)$,
so that
\be  \label{a5}
B_{11}(T) = \frac{M_{11}(T)}{M^{{\rm E}}_{11}(T_{\rm E})} =
\frac{\omega }{\omega_{\rm E}}
\ee
with $\omega_{\rm E}$ defined in (\ref{omega-Ensk}). This implies
that the reduced moments $B_{nm} (T)$ can all be expressed in
collisional averages, i.e.
\be \label{a6}
B_{nm}(T)  = \frac{\omega}{\omega_{\rm E}} \,
\frac{ \langle v_{12}^{n-1} |\cos\theta|^{m-1}
\rangle_{\hbox{\scriptsize coll}} }
{   \langle v_{12}^{n-1}
|\cos\theta|^{m-1}\rangle\eno_{\hbox{\scriptsize coll}} } .
\ee
The average $\langle \cdots \rangle\eno_{\hbox{\scriptsize coll}}$
is defined through (\ref{a1}) with
 $M_{nm}(T)$ replaced by $M^{{\rm E}}_{nm}(T_{\rm E})$, and
 calculated in (\ref{MEnm}). It
 represents the collisional average, evaluated with the Enskog
 factorization $f^{(2)} = \chi_{{\rm E}} f f
 $ {\it and also}
 taken at  the reference temperature $T_{{\rm E}}$.
 Note that the equality (\ref{a6}) consists of two factors,
$\omega$ and $\langle \cdots\rangle_{\hbox{\scriptsize coll}}$,
which are measured separately in event driven MD simulations.

We also observe that the equality $ \Gamma(T) = \Gamma_{\rm
E}(T_{\rm E})$, explained above (\ref{eq:ratio-gamma}), implies
that
\be \label{B33}
B_{33}(T) = \frac{M_{33}(T)}{M^{{\rm E}}_{33}(T_{\rm E})}=
\frac{\Gamma(T)}{\Gamma^{{\rm E}}(T_{\rm E})}= 1.
\ee
Furthermore we have for the excess pressure, $p^{{\rm ex}}(T)
\equiv p(T) - n T $,
\be \label{B22}
B_{22}(T)  =  \frac {p^{\rm ex}(T)/nT }{p_{\rm E}^{\rm ex}(T_{\rm
E})/n T_{\rm E}} =
\frac{\omega}{{\omega}_{\rm E} } \times
   \frac{\langle |v_{12} \cos\theta|\rangle_{\hbox{\scriptsize coll}}
}{\langle |v_{12} \cos\theta|\rangle^{\rm E}_{\hbox{\scriptsize
coll}} }
\ee
and for the dynamic pair correlation at contact $\chi^{(-)}$,
\be \label{B00}
B_{00}(T) = \frac{\chi^{(-)}}{\chi_{\rm E}} =\frac{\langle
|v_{12} \cos\theta|^{-1}
\rangle_{\hbox{\scriptsize coll}}}{\langle
|v_{12} \cos\theta|^{-1}\rangle^{\rm E}_ {\hbox{\scriptsize
coll}}}.
\ee
In the appendix we present a more complete set of relations for
the $B_{nm}$.

In the next section MD simulations will show that the predicted
deviation from a Maxwellian (see    (\ref{eq:expansion})) in the
(unconstrained) velocity distribution of a single particle can be
observed for small inelasticities. However {\em larger}
deviations are found between the observed constrained probability
distributions and averages, and the corresponding kinetic theory
predictions given by    (\ref{eq:averages}), based on molecular
chaos. Consequently the small corrections resulting from $a_2$ in
    (\ref{eq:expansion}) can be neglected in most cases.

%%%%%%%%%%%%%%%%%%%%%%%%%%%%%%%%%%%%%%%%%%%%%%%%%%%%%%%%%%%%%%%%%%%%

\subsection{Radial distributions}

The static or unconstrained  radial  distribution function in the
spatially homogeneous IHS fluid is defined as,
\be \label{eq:g-unc}
g(r)=\int \frac{{\rm d}\hat{\bsigma}}{{\Omega}_d} \int{\rm d}{\bf c}_1
{\rm d}{\bf c}_2  f^{(2)}({\bf c}_1,{\bf c}_2, r \hat{\bsigma}).
\ee
It may be averaged over all directions of ${\bf r}$ because of
statistical isotropy. The unconstrained radial distribution
function at contact is defined  as the extrapolation, $Y = g( r
\to \sigma + 0)$. By splitting the $\hsigma-$integration into a
precollision $ (\hg \cdot
\hsigma <0)$ and postcollision hemisphere $( \hg \cdot \hsigma > 0)$,
we obtain $Y$ as sum of two terms,
\be \label{sumY}
Y = \frac{1}{2} \left( Y^{(-)} + Y^{(+)}\right).
\ee
The definitions of $Y^{(-)},Y^{(+)}$ follow from
   (\ref{eq:g-unc}) by adding respectively factors
     $\Theta (-\hg \cdot \hsigma) $ and
 $\Theta (\hg \cdot \hsigma)$ under the integral sign
 in (\ref{eq:g-unc}).  The dummy integration
variables in $Y^{(+)}$ represent the postcollision velocities,
$({\bf c}^*_1,{\bf c}^*_2)$, corresponding to the precollision
ones, $({\bf c}_1,{\bf c}_2)$.

On the other hand, we have the dynamic precollision correlation
$\chi^{(-)}$, defined in (\ref{eq:g-min}), and a similar
postcollision one, $\chi^{(+)}$, defined by replacing $\Theta
(-\hg \cdot \hsigma)$ in (\ref{eq:g-min}) by $\Theta (\hg
\cdot
\hsigma)$. They are related by continuity of flux.
Because the incident flux of $({\bf c}_1{\bf c}_2)-$pairs just
before collision is equal to the scattered flux of $({\bf
c}^*_1{\bf c}^*_2)-$pairs just after collision, we have inside
dynamic averages the equality,
\be \label{fluxcont}
\Theta (-g_n) |g_n| f^{(2)}( {\bf c}_1{\bf c}_2, \bsigma)
{\rm d}{\bf c}_1 {\rm d}{\bf c}_2 {\rm d} \hsigma =
\Theta (g^*_n)|g^*_n| f^{(2)}( {\bf c}^*_1{\bf c}^*_2, \bsigma)
{\rm d}{\bf c}^*_1 {\rm d}{\bf c}^*_2 {\rm d} \hsigma ,
\ee
where $g_n = {\bf g} \cdot \hsigma = g \cos \theta $. The
reflection law, $g^*_n  =\alpha |g_n| $, for inelastic collisions
in combination with the continuity of the flux and
(\ref{eq:g-min}) yields at once,
\be  \label{unc-min}
\chi^{(+)} = (1/\alpha) \chi^{(-)}.
\ee
In principle, equations (\ref{sumY}) and (\ref{eq:g-min-coll})
provide two alternative routes to compute the precollisional pair
correlation at contact: the first one, denoted by $Y^{(-)}$, can
be implemented numerically as a static or unconstrained average,
namely by extrapolation to contact of the pair correlation
function for pairs aiming to collide. The second one, denoted by
$\chi^{(-)}$, can be computed as a dynamic collisional average,
calculated from $f^{(2)} ( {\bf c}_1, {\bf c}_2,
\bsigma)$ at contact. It is important to stress that the dynamic
$\chi^{(-)}$ is calculated as a time average over the {\it subset
of colliding pairs at contact}, and the static $Y^{(-)}$ as a
time average over {\it all pairs}, satisfying the relation, ${\bf
g}
\cdot \hsigma < 0$ and extrapolated to $r \to \sigma +0$, i.e.
calculated from $f^{(2)} ( {\bf c}_1, {\bf c}_2, {\bf r})$, where
the limit is taken after all integrations have been performed.
This may lead to different results, because the integrand contains
the the function $f^{(2)}$ which turns out to be singular
near $r=\sigma$ and
${\bf v}_{12}$ small (see discussion in subsection III C).

Physically, it is also clear why the averages in the NESS  need
not be the same. For instance, the relation (\ref{fluxcont}) may
not hold for the limiting $(r \to \sigma)$ values, $Y^{(-)}$ and
$Y^{(+)}$,  because non mean-field effects (in particular, the
'rotation-induced' recollisions discussed at the start of section
III, or noise induced recollisions, see below) may result in
differences between the two methods to evaluate $\chi^{(-)}$ and
$Y^{(-)}$. The reason is that the validity of (\ref{fluxcont}),
expressing flux continuity for the limiting values $(r \to \sigma
+0)$, is questionable in the presence of the external random
force. When the kicking frequency is much larger than the
collision frequency (situation considered here), a pair of
particles may indeed be put in contact under the action of the
random force only. We will investigate possible numerical
differences between $\chi^{(-)}$ and $Y^{(-)}$ in the next
section on MD simulations.

In paper I we have studied the long range spatial correlation
functions $G_{ab}(r)$ of the density field $n(\br,t)$ and the
flow field ${\bf u}(\br,t)$ in the NESS. These functions are
closely related to    (\ref{eq:g-unc}), i.e.
\ba \label{Gnn}
G_{nn}(r)&=&\frac{1}{n^2} \biggl\langle \sum_i \delta(\br_i-\br)[\sum_j
\delta(\br_j)-n]\biggr\rangle\nonumber\\
&=&\frac{1}{n}\delta(\br)+\left(g(r)-1\right),
\ea
and, in the notation of paper I,
\ba \label{Guu}
G_{{\bf u}{\bf u}}(r)&=&\frac{1}{n^2}\biggr\langle \sum_{i,j}
\bv_i\cdot\bv_j \delta(\br_i-\br) \delta(\br_j)\biggr\rangle\nonumber\\
&=&G_\parallel(r)+(d-1)G_\perp(r)\nonumber\\
&=&\textstyle{\frac{d}{2}} \frac{v_0^2}{n} \delta(\br)+ v_0^2
g(r) \langle{\bf c}_1\cdot{\bf c}_2\rangle (r),
\ea
where $\langle \cdots\rangle$ is an average over the $N$-particle
non-equilibrium steady state and the static velocity correlation,
$\langle{\bf c}_1\cdot{\bf c}_2\rangle (r)$, is defined  as,
\be \label{a21}
\langle {\bf c}_1\cdot{\bf c}_2\rangle (r) =
\int \frac{{\rm d}\hat{\bsigma}}{{\Omega}_d}
\int{\rm d}{\bf c}_1 {\rm d}{\bf c}_2
f^{(2)}({\bf c}_1,{\bf c}_2, r\hat{\bsigma}) ({\bf c}_1\cdot{\bf
c}_2) /g(r).
\ee
The correlation functions $G_{ab}(r)$ above are very long ranged,
decaying like $r^{2-d}$ for large distances. In the first part of
this subsection we have introduced the static correlations, $Y$,
$Y^{(\pm)}$, and the dynamic ones, $\chi^{(\pm)}$. In (\ref{a22})
and (\ref{a23}) we have done the same for the dynamic counter
parts $ \langle { \bf c}_1 \cdot {\bf c}_2\rangle_{\rm dyn}$ and
$\langle {\bf c}_1 \cdot {\bf c}_2\rangle^{(-)}_{\rm dyn}$ of the
static correlation $\langle {\bf c}_1 \cdot {\bf c}_2\rangle (r
\to \sigma) $, introduced in (\ref{Guu}).

In the next section the short range behavior of these functions
will be studied by MD simulations.

%%%%%%%%%%%%%%%%%%%%%%%%%%%%%%%%%%%%%%%%%%%%%%%%%%%%%%%%%%%%%%%%%
%%%%%%%%%%%%%%%%%%%%%%%%%%%%%%%%%%%%%%%%%%%%%%%%%%%%%%%%%%%%%%%%%%
\section{Simulation Results for the NESS}
To investigate the short scale
structure characterizing the NESS and the validity of molecular
chaos we will present in this section MD simulation results,
and compare these with our
theoretical predictions whenever possible.
The details of the simulations of the randomly
driven inelastic hard disk system have been reported elsewhere
\cite{pre}. We will work in the limit in which the kicking
frequency of the external random force is much larger than the
collision frequency. This is the limit in which the Fokker-Planck
term in (\ref{eq:fp}) models the random energy input through the
random kicks. The external random force will, in principle, have a
quantitative influence on the short range structure of the fluid.
There is only one important difference with respect to the
simulations carried out in
\cite{pre}. There,  the random rotation proposed in
\cite{deltour} was implemented to avoid inelastic collapse at
high inelasticities ($\alpha<0.5$). This procedure amounts to
rotating the relative velocity $\bf  \bg$ by a small random angle
after each collision. Consider the completely inelastic situation
$\alpha=0$ for the sake of the argument. After each collision,
the vector $ \bg$ lies exactly at the border of the
precollisional hemisphere ($\bg\!\cdot\!\hsigma=0 $), so that if
the aforementioned random angle has equi-probable positive and
negative values, the rotation procedure will lead to a
recollision with probability 1/2. This leads to a spurious
increase of the number of collisions by a factor
$\sum_{n=0}^\infty 1/2^n = 2$ (the recollision can itself induce
a recollision with probability 1/2 etc\ldots so that the
frequency of collision effectively doubles !). When $\alpha$ is
small but non vanishing, this effect is still present but weaker.
This is clearly an artificial violation of molecular chaos that
has been discarded in the present work: for $\alpha<0.5$, we have
also implemented the rotation procedure, but if a small rotation
leads to a recollision, a new angle is drawn until the pair
separates. In this way, we reduce an important source of
correlations (the effect is dramatic on all the low order moments
$B_{nm}$, not only on the collision frequency; in particular, the
moments with $n\leq 1$ that correspond to collisional averages of
negative powers of $g$, are strongly biased toward bigger values
if the ``rotation-induced recollisions'' are present). After
applying this new rule, we are then left only with correlations
induced by the hard sphere dynamics plus the ones induced by the
noise itself (see below).

\label{sec:md}
\subsection{Cumulants}
First we focus on the single particle velocity distribution
function $ f$ averaged over all particles, which deviates from a
Maxwellian distribution due to the inelasticity of the collisions.
In the previous section we presented predictions for these
deviations, assuming molecular chaos. The resulting expression
given by     (\ref{eq:a2}) for the fourth cumulant $a_2$ of the
distribution as well as the prediction for its overpopulated tail
are in perfect agreement with 3D Direct Simulation Monte Carlo
(DSMC) results over the whole region of inelasticities
\cite{santos}.
As DSMC itself invokes molecular chaos, this observation merely
justifies the approximations made in the analytic calculation.
Information about the validity of molecular chaos can only be
obtained from a comparison with molecular dynamics simulations,
and Fig. \ref{fig:a2} shows this comparison for the fourth
cumulant $a_2$ as a function of the coefficient of restitution
$\alpha$ in 2D. The simulation results are in agreement with
(\ref{eq:a2}) for small inelasticity, but start to deviate
significantly from the theoretical prediction at $\alpha=0.6$.
These deviations, together with the perfect agreement between the
theoretical prediction and DSMC results, provide direct evidence
for the breakdown of molecular chaos for $\alpha \lesssim 0.6$.
The theoretical result (\ref{eq:a2}) is independent of the
density. As $a_2$ represents only a small correction in
(\ref{eq:expansion}), one needs a large number of collisions and
a large number of particles to reach sufficient statistical
accuracy. So, high densities, for which one can use linked lists
\cite{linked}, are well suited.
The data in Fig. \ref{fig:a2} are typically obtained at high
packing fractions ($\phi=0.63$ at $\alpha=0.92$ and $0.7$, and
$\phi=0.55$ at $\alpha=0.6$) and $N=10201$ particles. At low
densities and weaker inelasticities ($\alpha\gtrsim 0.6$) we are
unable to collect enough statistics to measure the small
correction, represented by $a_2$. Simulations at higher
inelasticities did not show any density dependence of $a_2$ in
the range $0.2 \lesssim \phi \lesssim 0.6$, suggesting that the
cumulant expression     (\ref{eq:cumul}), obtained from the
Enskog-Boltzmann equation also applies at liquid densities.
 Similar results as those displayed in Fig. \ref{fig:a2} have been
observed for the 3-dimensional version of the present model \cite{moon}.

%%%%%%%%%%%%%%%%%%%%%%%%%%%%%%%%%%%%%%%%%%%%%%%%%%%%%%%%%%%%%%%%%%%%%%%%%%%

\subsection{Radial distribution function}
\label{subsec:contact}
Next, we present results for the static or unconstrained radial
distribution function $g(r)$, and in particular  its extrapolated
values at contact $Y$, $Y^{(-)}$, $Y^{(+)}$. Here $g(r)$ is
essentially the density-density correlation function, whose long
range behavior was studied in Ref.\ \cite{pre}.

Figure \ref{fig:gr} shows the measured values of $g(r)$ for short
distances and packing fraction $\phi=0.2$, at different
inelasticities. At small inelasticities ($\alpha\simeq 0.9$),
$g(r)$ resembles the radial distribution function for elastic
hard disks (EHD). At higher inelasticities, deviations start to
appear: the first and second maximum  in the measured $g(r)$ are
enhanced with respect to their EHD values at the same density.
Moreover, the functional shape also deviates from the
corresponding pair distribution of EHD at an appropriately chosen
{\em higher} density; e.g.\ if this density is chosen such that
the value of the second maximum of the pair distribution of EHD
coincides with the simulation result for IHD, the observed value
at contact would still be underestimated by the EHD pair
distribution.

It seems worthwhile to compare these results with existing
experiments on granular fluids in which the pair distribution
$g(r)$ has been measured. In the experiment of Ref.\
\cite{urbach} on a vertically vibrated thin granular layer,
$g(r)$ has been measured at
 $\phi=0.46$.
In the fluidized (`gas-like') phase, it follows the equilibrium
result for elastic hard disks almost identically. This result may
be compared to our simulations for a randomly driven fluid of
inelastic disks at $\alpha=0.9$, corresponding to the value for
stainless steel balls used in the experiment. It would be of
interest to measure experimentally how $g(r)$ in the fluidized
phase depends on the inelasticity, and see if a behavior similar
to that of Fig.
\ref{fig:gr} is observed. It is also interesting to note that the
pair correlation function $g(r)$ in a non-Brownian suspension of
spherical particles, fluidized between two vertical parallel
plates, shows an enhanced value at contact as well \cite{martin}.

In Fig. \ref {fig:grcontact}a we show the value at contact, $Y$,
 obtained by extrapolation from $g(r)$ at
$\phi=0.05$, together with the extrapolated values for approaching
and receding pairs, $Y^{(-)}$ and $Y^{(+)}$ respectively. For
$\alpha \gtrsim 0.8$ no significant deviations are found from the
Verlet-Levesque value $\chi_{\rm E}= 1.084 $ for elastic hard
disks at the same density. More surprising is the value of
$g(r=\sigma)$ at large inelasticities, reaching a value around 40
for $\alpha
\to 0$. This property, combined with the observation that the
first and second maximum in $g(r)$ are shifted to smaller
$r$-values, and are larger (up to 20\% at small $\alpha$) than
the corresponding hard disks values, may be interpreted as a
tendency to cluster, i.e. to stay in continuously rearranging
configurations with large density inhomogeneities. We return to
this point in subsection III G.

Figure \ref{fig:grcontact} also shows the dynamic correlation, $\chi^{(-)}=
\chi_{{\rm E}} B_{00}$, measured as a collisional average.
Fig. \ref{fig:grcontact}b compares the static ratio
$Y^{(-)}/Y^{(+)}$ with the dynamic one, $\chi^{(-)}/\chi^{(+)}
=\alpha$, and also shows the ratio $Y^{(-)}/\chi^{(-)}$. The
plots clearly show that the dynamic and static correlations,
$\chi^{(\pm)}$ and $Y^{(\pm)}$, are different. For  $\alpha
\lesssim 0.5$  the differences are large,  and for $\alpha
\gtrsim 0.6$ both functions are about equal.
For the case of a freely evolving IHS fluid, Soto and Mareschal \cite{SM}
have recently observed a similar behavior, and explained it in
terms of the effect of the increase of grazing collisions on the
effective $\chi^{(-)}$. In the randomly driven IHD fluid the same
effects are present.  
All correlations, $\chi^{(-)}$, $Y^{(\pm)}$ are large, especially at small
$\alpha$. This is caused by the divergence of 
$f^{(2)}({\bf c}_1,{\bf c}_1, {\bf \sigma})$ at
small $g$ and small $\cos \theta$, which corresponds to grazing
collisions and will be further discussed in the next subsection. 
As a result of noise-induced recollisions, collisions with small $g$ 
and small $\cos \theta$ are oversampled; consequently,  a dynamical average 
involving negative powers of $g \cos \theta$ such as $\chi^{(-)}$
is expected to be larger than its static counterpart $Y^{(-)}$.
This feature can be observed in Fig. \ref{fig:grcontact}b.
Moreover, in the absence of recollisions, we would expect 
$Y^{(-)} = \alpha Y^{(+)}$ as a result of plain hard sphere dynamics. 
However, in the heated system, the flux continuity, as expressed in
Eq.   (\ref{fluxcont}), is no longer satisfied for the extrapolated
$Y$'s; some pairs are put in pre-collision configuration under
the action of the random force, which leads to $Y^{(-)} > \alpha Y^{(+)}$.
The breakdown of $Y^{(-)} = \alpha Y^{(+)}$
signals the inelasticity  beyond which noise-induced correlations become relevant.
It is furthermore possible to consider situations where the recollisions dominate
the dynamics, e.g. at small $\alpha$ by allowing rotation-induced recollisions.
In this extreme limit, we expect $Y^{(-)} \simeq  Y^{(+)}$, the pairs 
being put in pre- or post-collision configuration essentially at random.
%This accounts for the increase of $Y^{(-)} / Y^{(+)}$ as $\alpha$ decreases
%and becomes smaller than 0.3 in Fig. \ref{fig:grcontact}b. 
In the same limit, the population of colliding pairs with 
small $g$ and small $\cos \theta$ is enhanced, leading to a more pronounced
discrepancy between dynamic and static averages (i.e. a much smaller ratio 
$Y^{(-)} / \chi^{(-)}$ than observed in Fig. \ref{fig:grcontact}b).

%%%%%%%%%%%%%%%%%%%%%%%%%%%%%%%%%%%%%%%%%%%%%%%%%%%%%%%%%%%%%%%%%%%%
\subsection{Equation of state. Molecular chaos breakdown}

\begin{center}
\begin{figure}[h]
\centerline{\hspace{-8cm}\psfig{figure=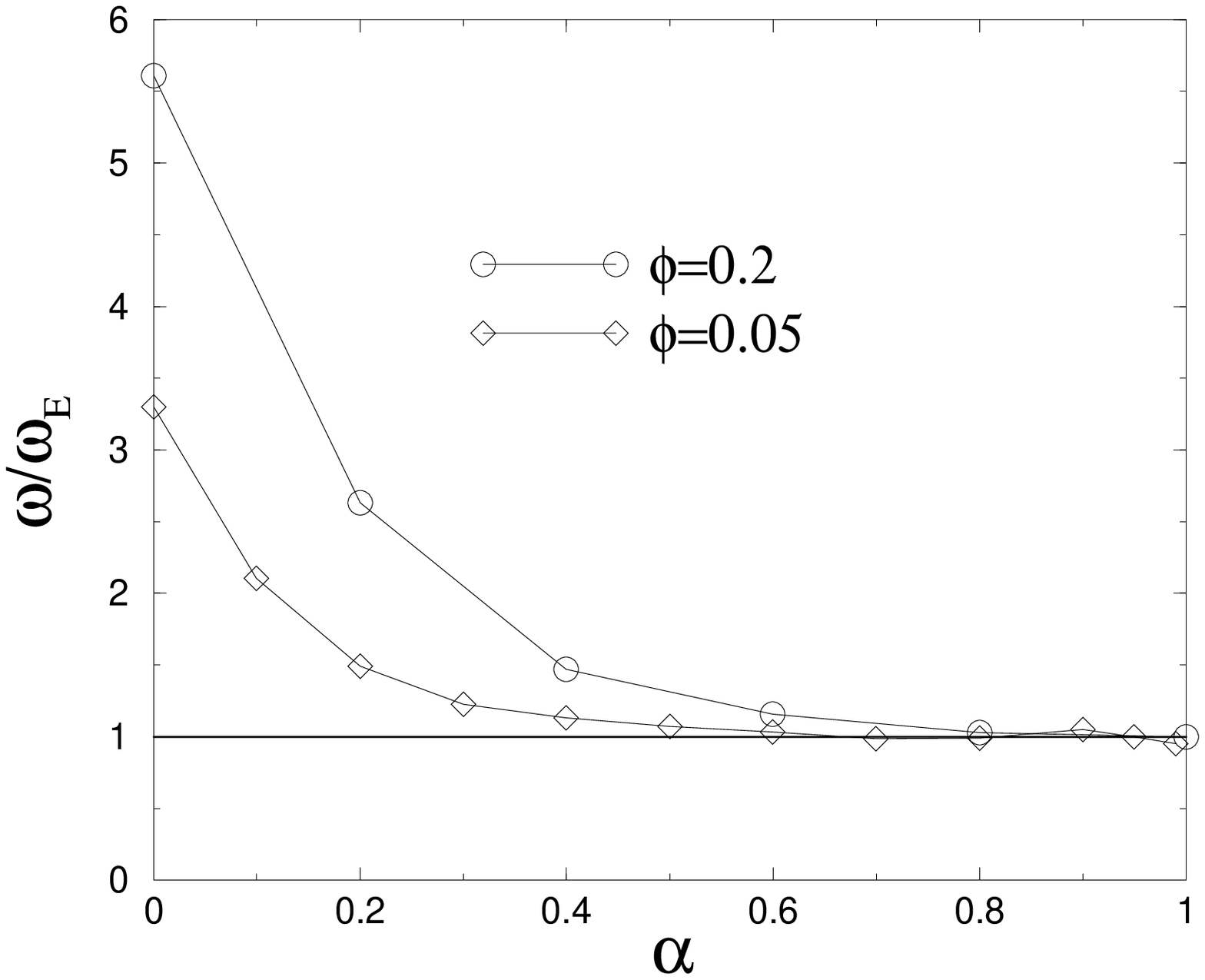,height=6.5cm}}
\vspace{-6.5cm}
\centerline{\hspace{8cm}\psfig{figure=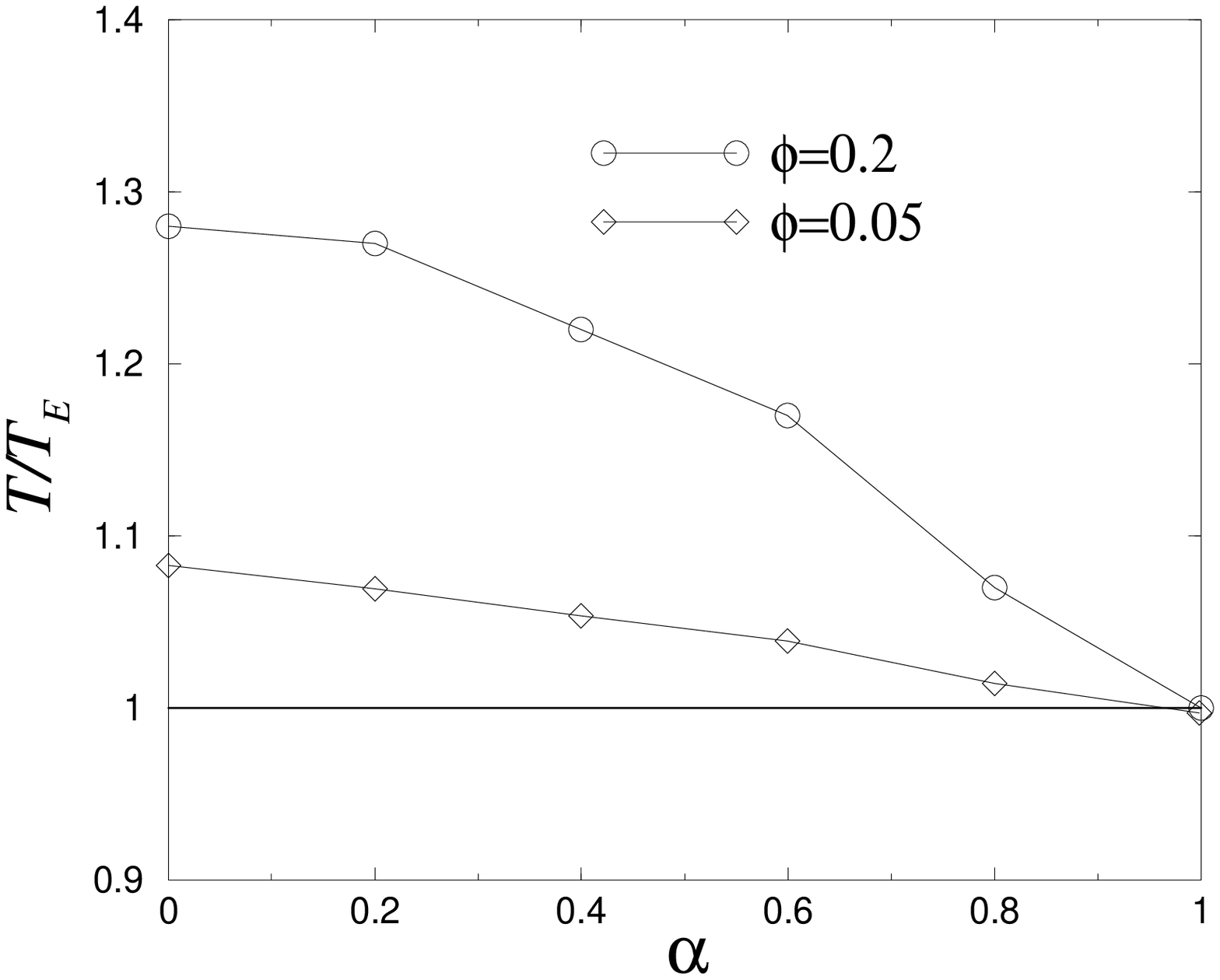,height=6.5cm}}
\caption{
\label{fig:temp_omega}}
\end{figure}
\end{center}

\begin{center}
\begin{figure}[h]
\vspace{1.5cm}
\epsfig{figure=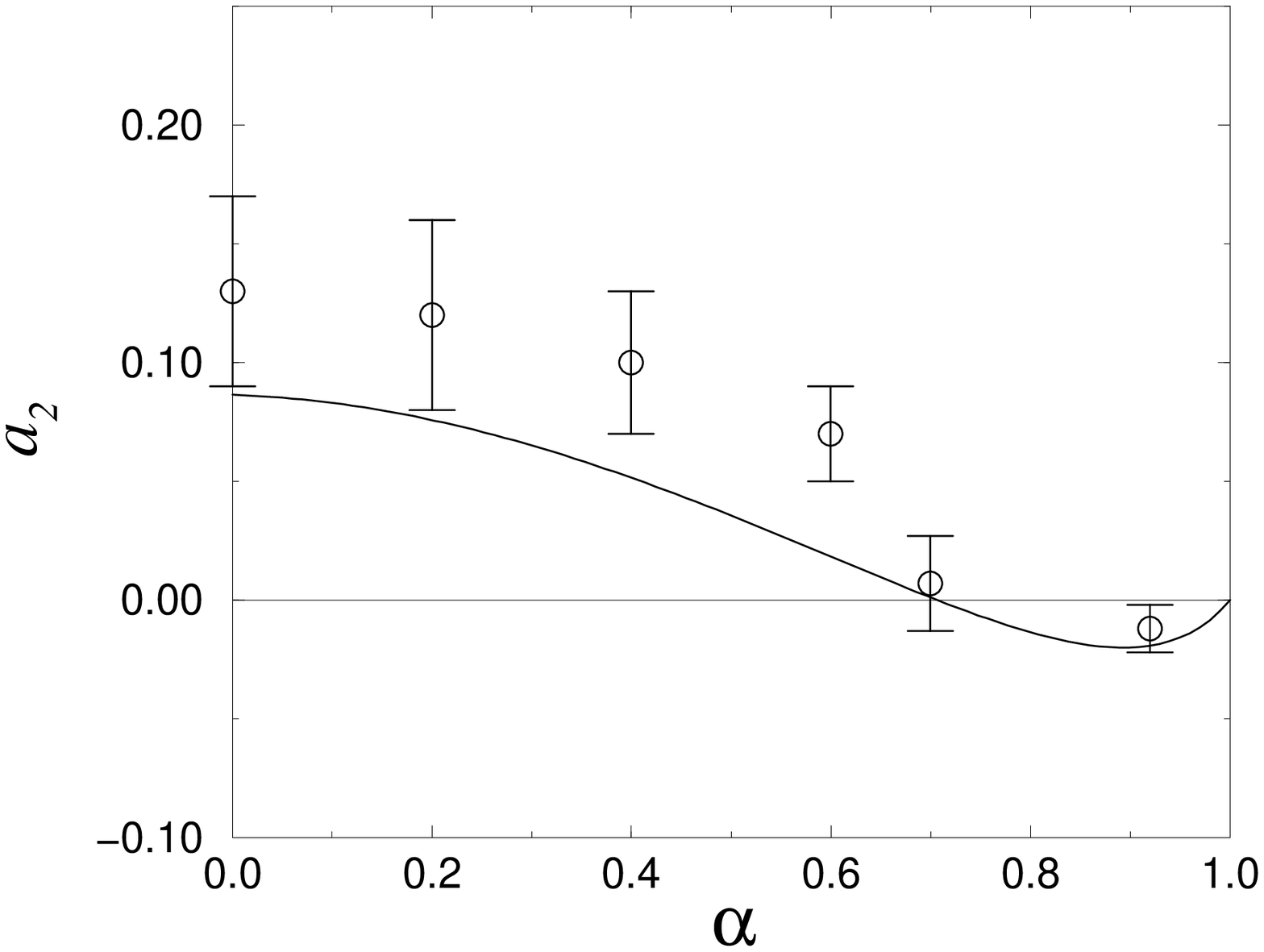,width=13cm,angle=0}
\caption{
\label{fig:a2}}
\end{figure}
\end{center}

\begin{center}
\begin{figure}[h]
\vspace{-0.5cm}
\epsfig{figure=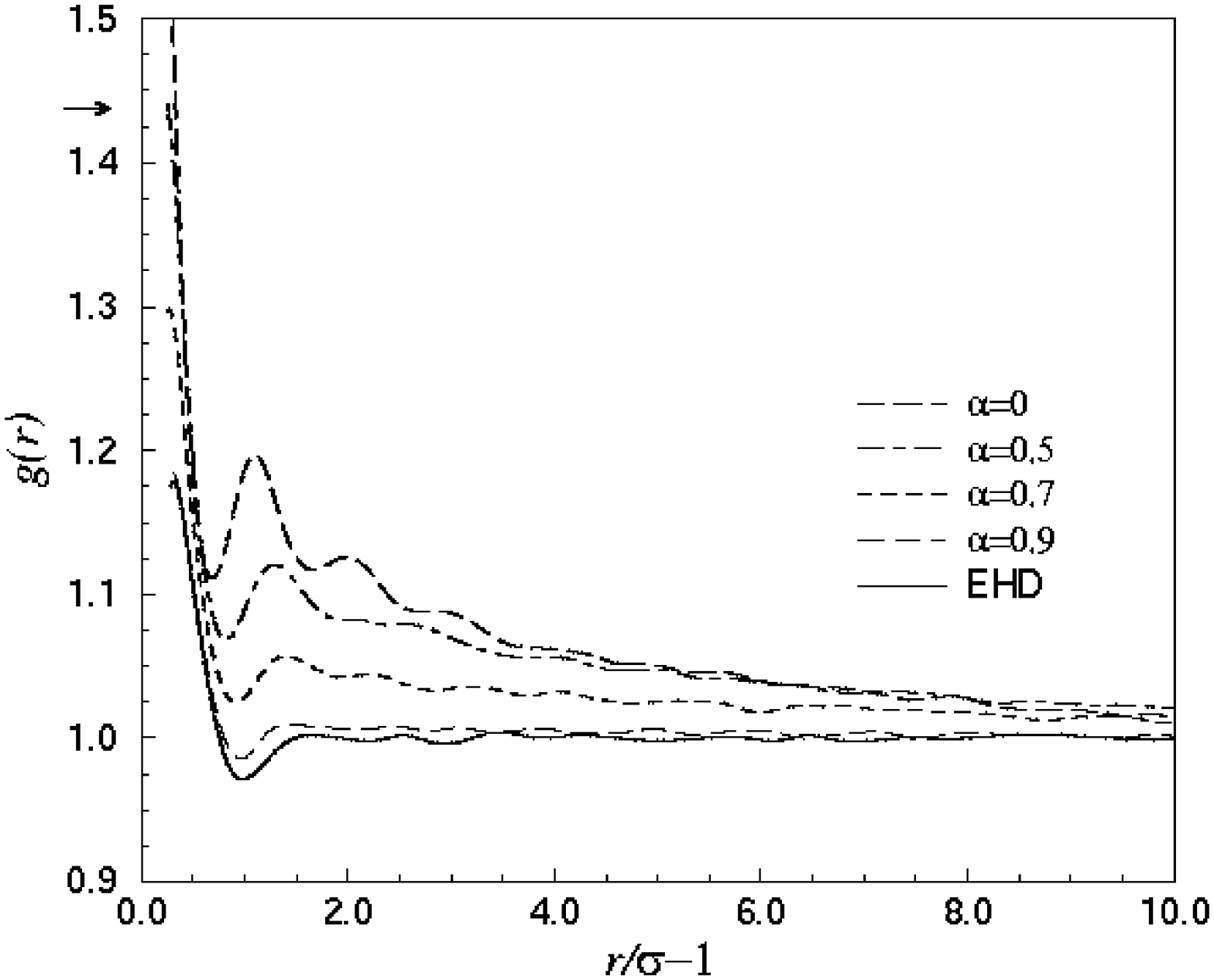,width=13cm,angle=0}
\caption{
\label{fig:gr}}
\end{figure}
\end{center}

\begin{center}
\begin{figure}[h]
\vspace{1.5cm}
\epsfig{figure=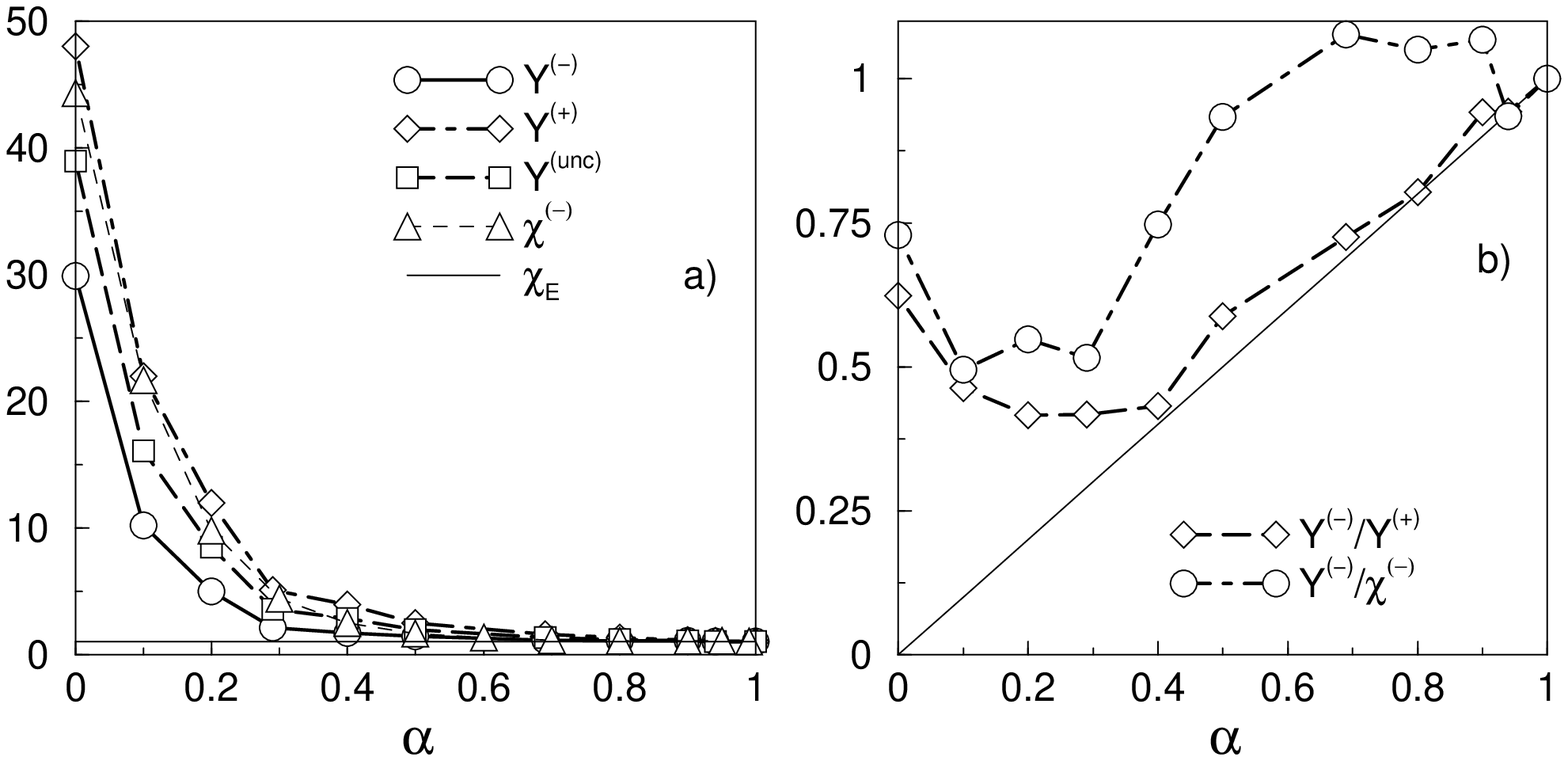,width=13cm,angle=0}
\caption{
\label{fig:grcontact}}
\end{figure}
\end{center}

To what extent does the extrapolated  static radial distribution
function, $Y = g(r \to \sigma)$, describe the nontrivial
dependence in the NESS of collision frequency in
(\ref{eq:omegadef}), collisional damping in (\ref{eq:gammadef})
and pressure in (\ref{eq:press}) on the inelasticity? If
molecular chaos holds, the latter quantities depend, according to
Eqs.\ (\ref{eq:mc}), (\ref{eq:pressmc}), (\ref{eq:omega-Ensk}) and
(\ref{eq:gamma-Ensk}), on the precollisional pair function at
contact, $\chi^{(-)}$, where the particles are aiming to collide.
This function differs from the extrapolated static   $Y^{(-)}$ at
high inelasticities (see Fig. \ref{fig:grcontact}). Consider first
the  collision frequency in the molecular chaos approximation,
$\omega_{{\rm mc}} =\chi \omega_0(T)$ above (\ref{eq:omega-Ensk}),
with $\chi = \chi^{(-)}$ the dynamic correlation, i.e.
\be
\label{w-appr-dyn}
\frac{\omega_{{\rm mc}} (T)}{\omega_{{\rm E}}} =
\frac{\chi^{(-)}}{\chi_{{\rm E}}}
\frac{v_0}{v_{{\rm E}}} = B_{00} \sqrt{ \frac{T}{T_{{\rm E}}}},
\ee
where we have used Eqs.(\ref{eq:omega-Ensk}) and (\ref{B00}).
This is an extremely poor approximation, as can be seen from
Fig. \ref{fig:temp_omega}, which shows that the measured value
$\omega/\omega_{{\rm E}}$ approaches 5.6 as $\alpha \to 0$,
whereas $B_{00}$ is
essentially divergent. Next we replace $\chi^{(-)}$ in
(\ref{w-appr-dyn}) by its static counterpart $Y^{(-)}$, shown in
Fig. \ref{fig:grcontact}. This yields
$\omega_{{\rm stat}}(T)/\omega_{{\rm E}} =
Y^{(-)} v_0/\chi^{(-)} v_{{\rm E}}$. Its limiting value for
$\alpha \to 0$ is about a factor 3 too large when compared to
$\omega/\omega_{{\rm E}}$. We conclude that all mean field
approximations for the collision frequency, including the Enskog
approximation $\omega_{{\rm E}}(T)/\omega_{{\rm E}} =
\sqrt{T/T_{{\rm E}}}$, break down for $\alpha <0.6$.

\begin{center}
\begin{figure}[h]
\vspace{-0.5cm}
\epsfig{figure=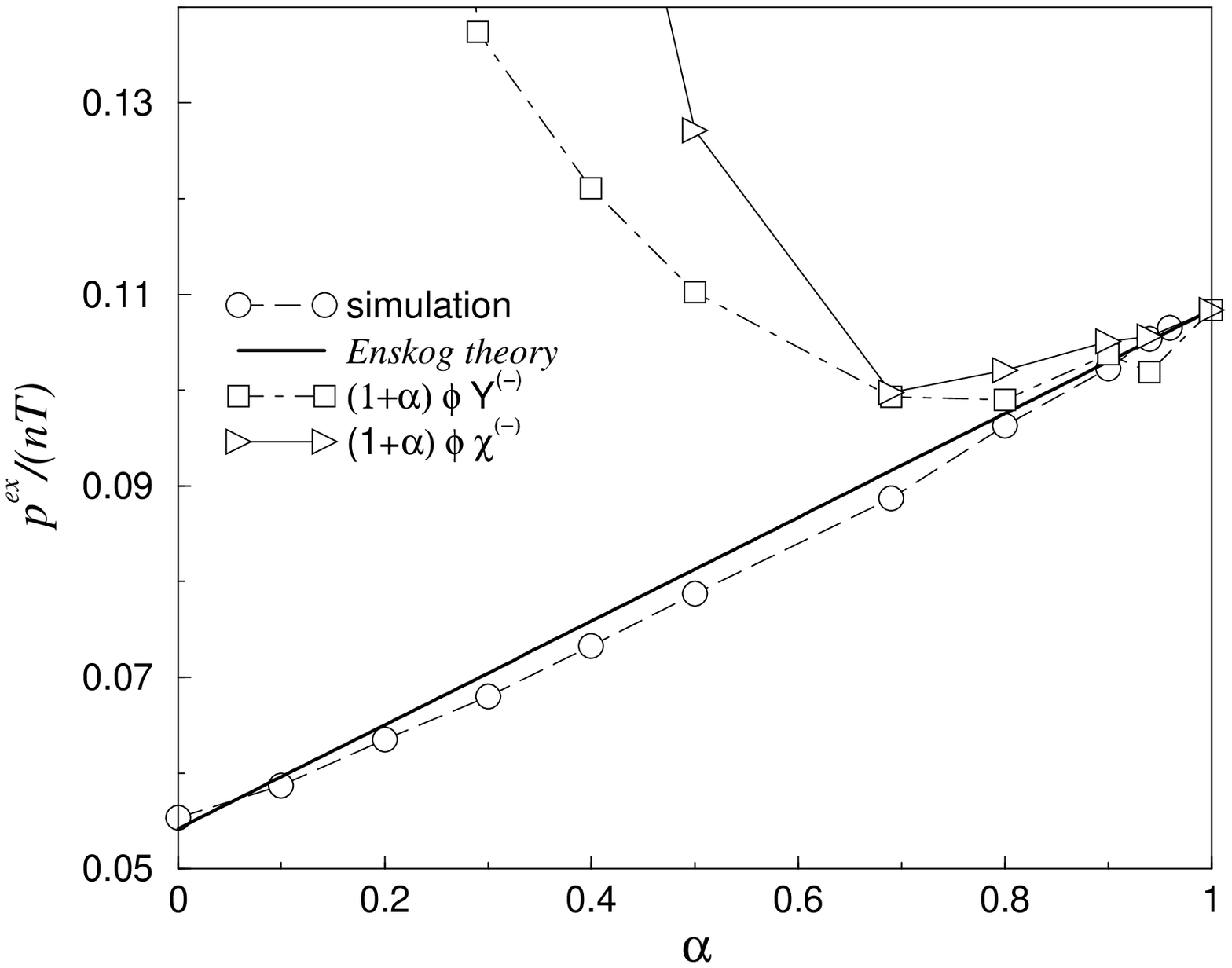,width=12cm,angle=0}
\caption{
\label{fig:virial}}
\end{figure}
\end{center}

Fig. \ref{fig:virial} shows the pressure of the IHD fluid,
compared with the molecular chaos prediction given by
(\ref{eq:pressmc}), taking for $\chi$ either the Enskog
approximation $\chi_{\rm E}$ in (\ref{eq:vlcs}), or the simulated
$Y^{(-)}$, or  $\chi^{(-)}$. The Enskog approximation, accounting
for the short range geometric exclusion effects in the
precollision state, gives a reasonable description of $p(T)$ for
all $\alpha$, while both the  static $Y^{(-)}$ and the dynamic
$\chi^{(-)}$ give an extremely poor description except for
$\alpha > 0.8$.

Consistent with this conclusion is the good estimate for the
temperature $T_{{\rm E}}$ in the NESS, obtained by balancing the
energy dissipation rate $\Gamma_{{\rm E}}(T_{{\rm E}})$ in
(\ref{T-Ensk-implicit}) with the energy input from the random
force, as shown in Fig. \ref{fig:temp_omega}. Moreover the
collisional energy loss, $\Gamma(T)/\Gamma_{{\rm E}}(T)=(T_{{\rm
E}}/T)^{3/2}$ in (\ref{eq:ratio-gamma}), is in agreement with MD
simulations over the whole $\alpha$- interval within $30 \%$. All
other mean field approximations with $\omega_{{\rm E}}$ replaced
by $\omega_{{\rm mc}}(T)$  or $\omega_{{\rm stat}}(T)$  give very
poor results for $\Gamma (T) =m \xi_0^2$.

\begin{center}
\begin{figure}[h]
\vspace{-0.5cm}
\epsfig{figure=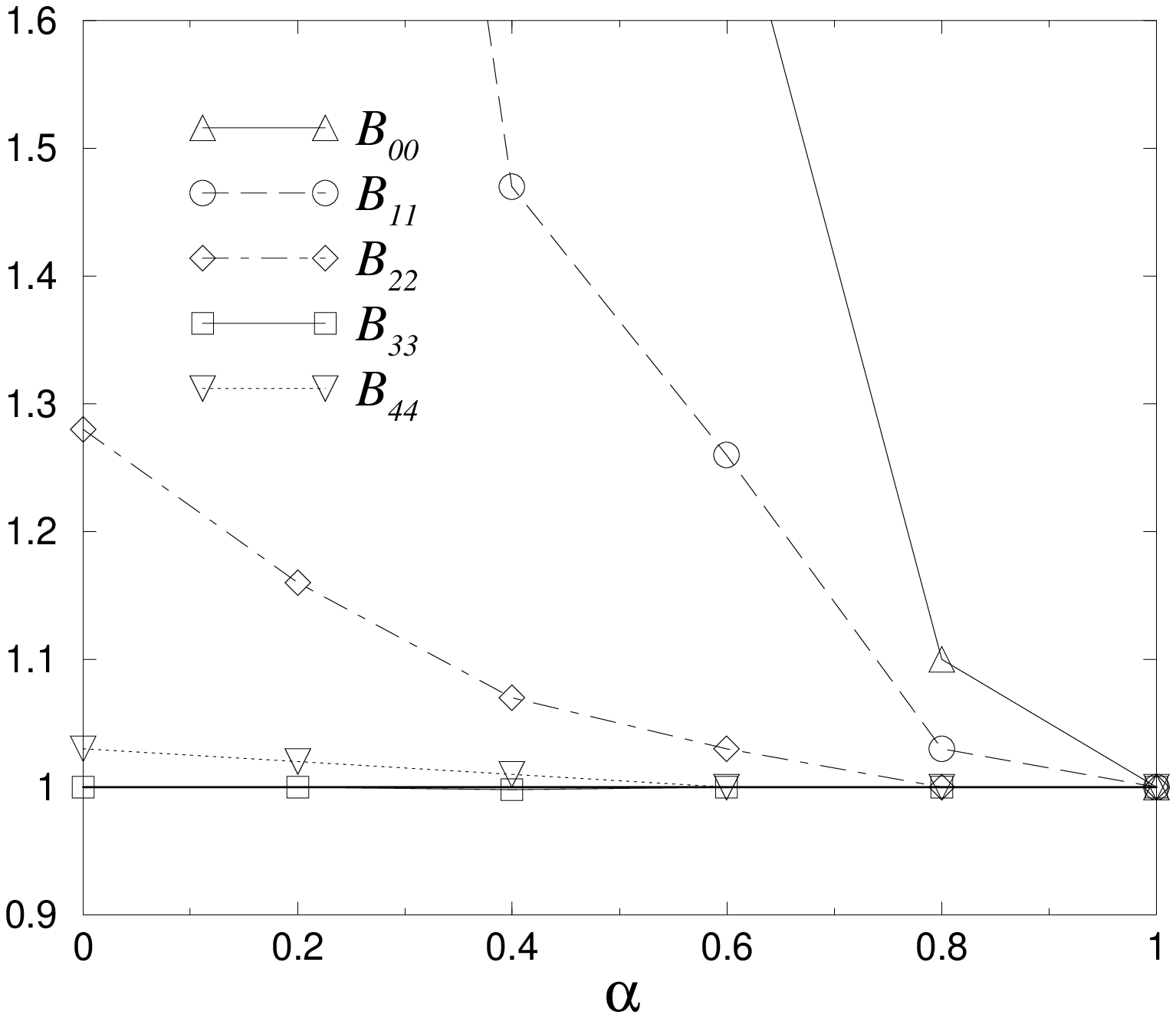,width=11cm,angle=0}
\caption{
\label{fig:Bnn}}
\end{figure}
\end{center}

How can these paradoxical results be reconciled? Let us  compare
the individual definitions of $\chi^{(-)},\omega, p$ and
$\Gamma$, which all contain factors $|{\bf g}\cdot {\bf \sigma}
|^n f^{(2)}({\bf c_1}, {\bf c_2}, \bsigma)$ with $n=0,1,2,3$. To
find a possible explanation  of these paradoxical results, we test
the following scenario: {\it the molecular chaos assumption
(\ref{eq:mc}) only  breaks down at very small relative velocities
${\bf g}$, and more precisely, at very small  $g_n= {\bf g}\cdot
{\hat{\bf \sigma}}= g \cos \theta $, which is the component of ${\bf
g}$, parallel to the line of centers of the colliding particles}
(physical arguments for this scenario will be offered in
subsection
\ref{sec:recoll} where we discuss the noise induced
recollisions). On the basis of this scenario the singularity in
$f^{(2)}$ at small ${\bf g}$ makes the dynamic correlation
$\chi^{(-)} = B_{00} \chi_{{\rm E}}$ ( shown in Fig. \ref{fig:Bnn}) very
much larger than $\chi_{{\rm E}}$, essentially divergent as
$\alpha \to 0$. In calculating the collisional frequency from
(\ref{eq:omegadef}) the extra factor $g_n$ in the integrand makes
the small $g_n-$singularity integrable, giving a {\it finite}
correction to the Enskog collision frequency, also for $\alpha
\to 0$ (see Fig. \ref {fig:temp_omega}). The contributions of the small
$g_n-$singularity in $p(T)$ and $\Gamma(T)$ are essentially
suppressed by extra factors of $|{\bf g}\cdot \hat{{\bf \sigma}}
|^n $.

\begin{center}
\begin{figure}[h]
\centerline{\hspace{-8cm}\psfig{figure=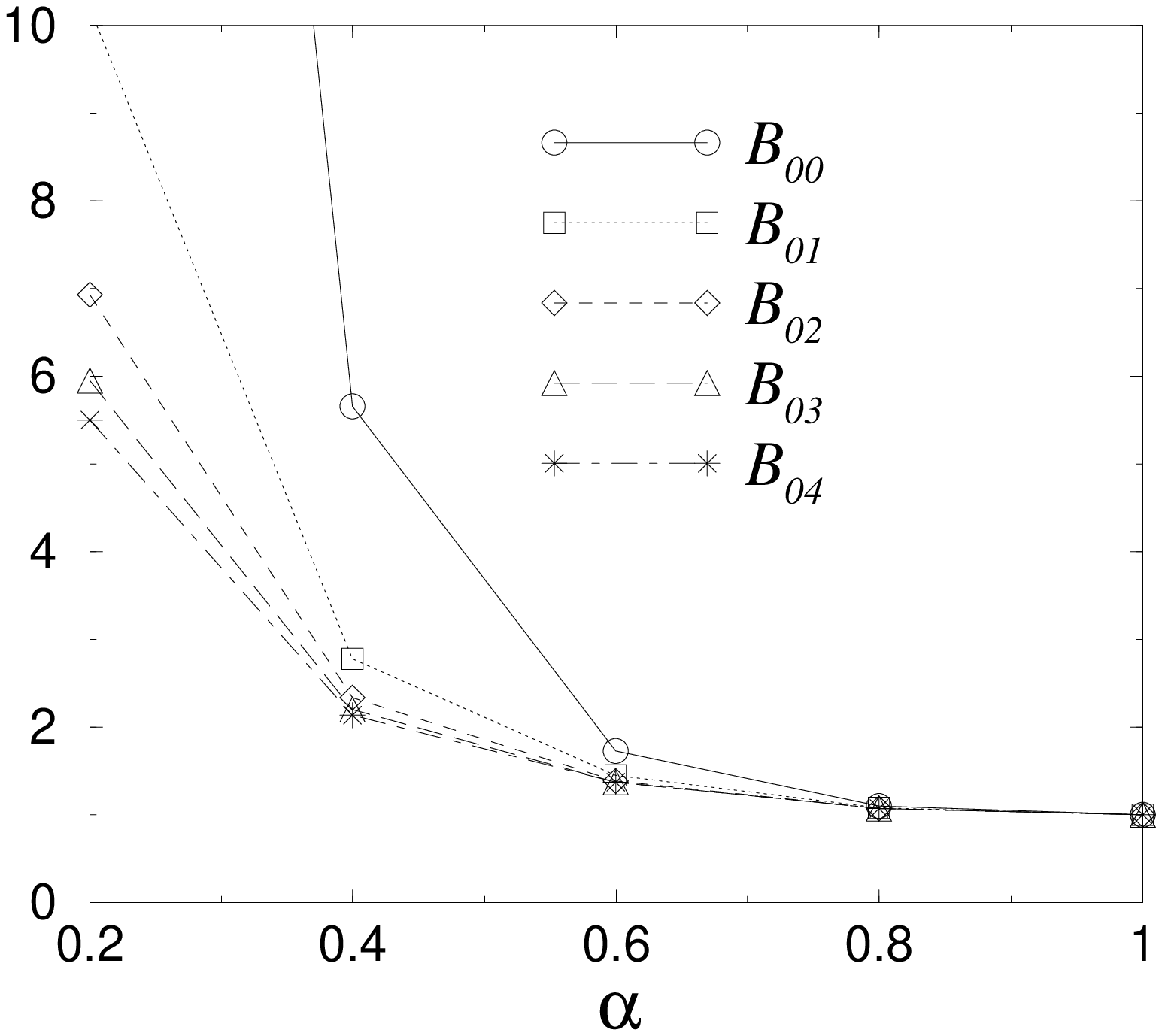,height=6.cm}}
\vspace{-6.cm}
\centerline{\hspace{8cm}\psfig{figure=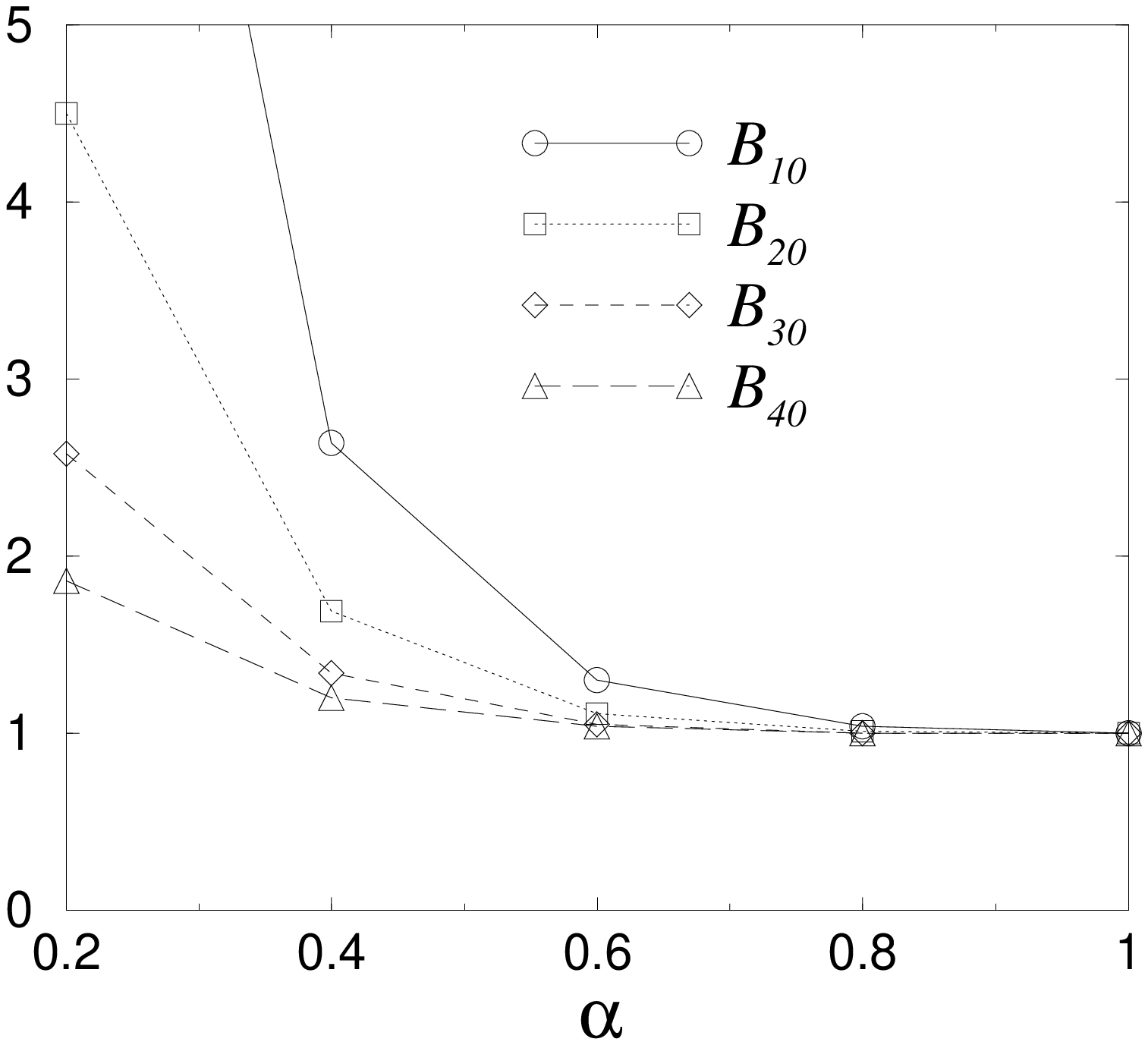,height=6.cm}}
\caption{
\label{fig:bnm}}
\end{figure}
\end{center}

This possibility has been analyzed systematically by measuring the
behavior of the moments $B_{nm}(T)$, which are useful tools to
investigate the breakdown of the molecular chaos assumption. We
have made $n$ and/or $m$ small in order to analyze the nature of
the singularities in $f^{(2)}$ near small relative velocities and
near grazing collisions, as displayed in Fig. \ref{fig:bnm}. All
deviations of these quantities from unity give a quantitative
measure for the violation of the molecular chaos assumption. In
the elastic limit we have carefully checked for a large number of
cases that the reduced moments $B_{nm}$ tend to 1.
Fig. \ref{fig:Bnn} shows the values of different moments
$B_{nn}(T)$, and one can clearly see how the deviations from the
elastic limit rapidly decrease as $n$ increases to $n=3$, after
which they start to increase slowly. For larger $n$-values, 
the moments are reasonably close to unity, but statistical inaccuracy
precludes any definite conclusion about the large $n$ behaviour.

Further evidence for the above scenario is shown in Fig. \ref{fig:bnm},
where we display two sequences of moments $B_{nm}$. To draw some
further conclusions from Figs. \ref{fig:Bnn} and
\ref{fig:bnm} we note that the integrands in $B_{nn},B_{0n}$ and
$ B_{n0}$, as defined in (\ref{Mnm}), contain apart from
$f^{(2)}$ respectively the factors $g^n |\cos \theta|^n,|\cos
\theta|^n, g^n$. The reduced moments $B_{01}$ and $B_{10}$
contain again very large contributions from the divergence of
$f^{(2)}$ near vanishing $g_n= g \cos \theta$. Fig. \ref{fig:bnm}
suggests that the presence of equal powers of $g$ and $\cos
\theta$ in $B_{nn}$ simultaneously suppresses the large
contributions from the singularities at $g=0$ and $\cos \theta
=0$.

We conclude that the numerical results, displayed in Figs.
\ref{fig:Bnn} and \ref{fig:bnm}, give support to the previous
scenario, showing that molecular chaos breaks down only in a very
small portion of the phase space, around $g_n= {\bf g \cdot
\hat{\sigma}}= g \cos \theta=0$. The size of this 'pocket'
in phase space increases as $\alpha$ decreases. Therefore, only
those collisional quantities that contain low powers of $g$ and
$\cos \theta$ (such as $\chi^{(-)}$ and $\omega$) will be very
sensitive to this breakdown as the inelasticity increases, while
physical quantities involving higher powers of $g$ and $\cos
\theta$, such as the temperature, pressure or energy dissipation
will be well approximated by their molecular chaos counterparts.

%Interestingly enough, by increasing $n>1$ and $m>1$, the
%coefficients $B_{nm}$ converge to 1 irrespective of $\alpha$ or
%the density, which is an indication that the two-body
%distribution function factorizes in the form
%\be
%\hat{f}^{(2)}\left( {\bf  v}_1,  {\bf v}_2, {\bf \sigma}| T\right)\, = \,
%\chi_{\rm E}\,f^{(1)}\left({\bf  v}_1|T\right)\,
%f^{(1)}\left({\bf  v}_2|T\right)
%\label{eq:factorization}
%\ee
%except for vanishing $|\bf  v_{12}|$ and $\cos \theta$ where $f^{(2)}$
%diverge. In Eq. (\ref{eq:factorization}), we made explicitly appear the
%temperature dependence; given that the measured $T$ differs from the
%Enskog predicted $T_{\rm E}$ ($T>T_{\rm E}$. see
%Fig \ref{fig:temp_omega}) systematically,
%the factorization property
%given by     (\ref{eq:factorization}) is then remarkable. It should
%however be remembered that the constraint concerning the damping
%rates implies $B_{33}=1$ identically (see
%(\ref{eq:constraint})). Moreover, we have restricted the analysis
%to moments of order $m\leq 4$ and $n\leq 4$, that are expected to
%be close to $B_{33}$ (except of course at low orders). We then
%emphasize that higher orders moments, not computed here, might
%deviate significantly from 1, which would invalidate Eq.
%(\ref{eq:factorization}).
%%%%%%%%%%%%%%%%%%%%%%%%%%%%%%%%%%%%%%%%%%%%

%%%%%%%%%%%%%%%%%%%%%%%%%%%%%%%%%%%%%%%%%%%%%%%%%%%%%%%%%%%%%

\subsection{Velocity correlations at contact}

In the previous subsection we have considered the pair
distribution function $ f^{(2)}({\bf c_1}, {\bf c_2}, {\bf
\sigma})$ in the precollision state, and have examined how
molecular chaos is broken down, and which physical quantities are
most sensitive to it. Now we will analyze the effect of the
breakdown of molecular chaos on collisional statistics.

\begin{center}
\begin{figure}[h]
\vspace{-1.5cm}
\epsfig{figure=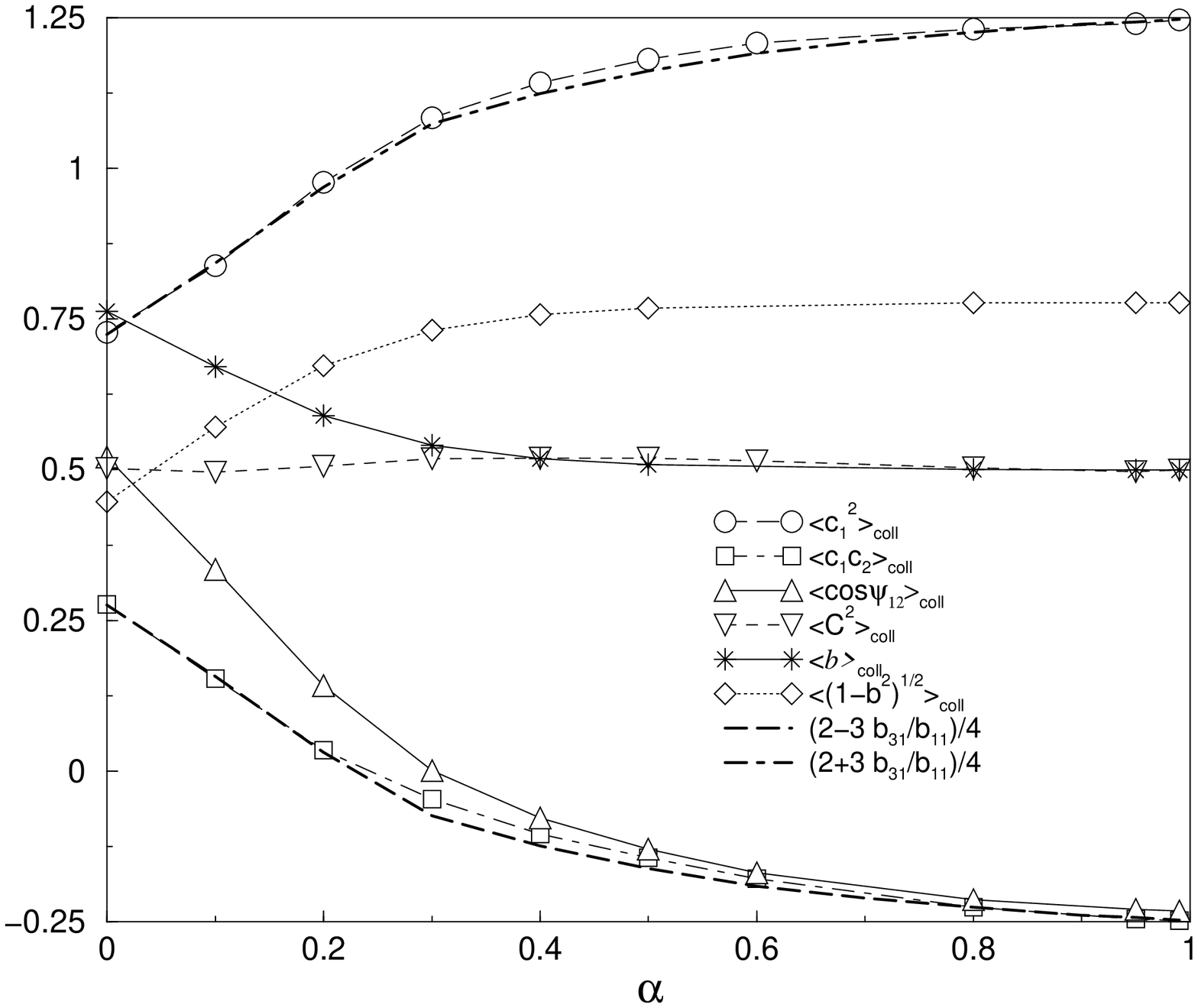,width=6.5cm,angle=-90}
\caption{
\label{fig:coll1}}
\end{figure}
\end{center}

\begin{center}
\begin{figure}[h]
\vspace{-0.2cm}
\epsfig{figure=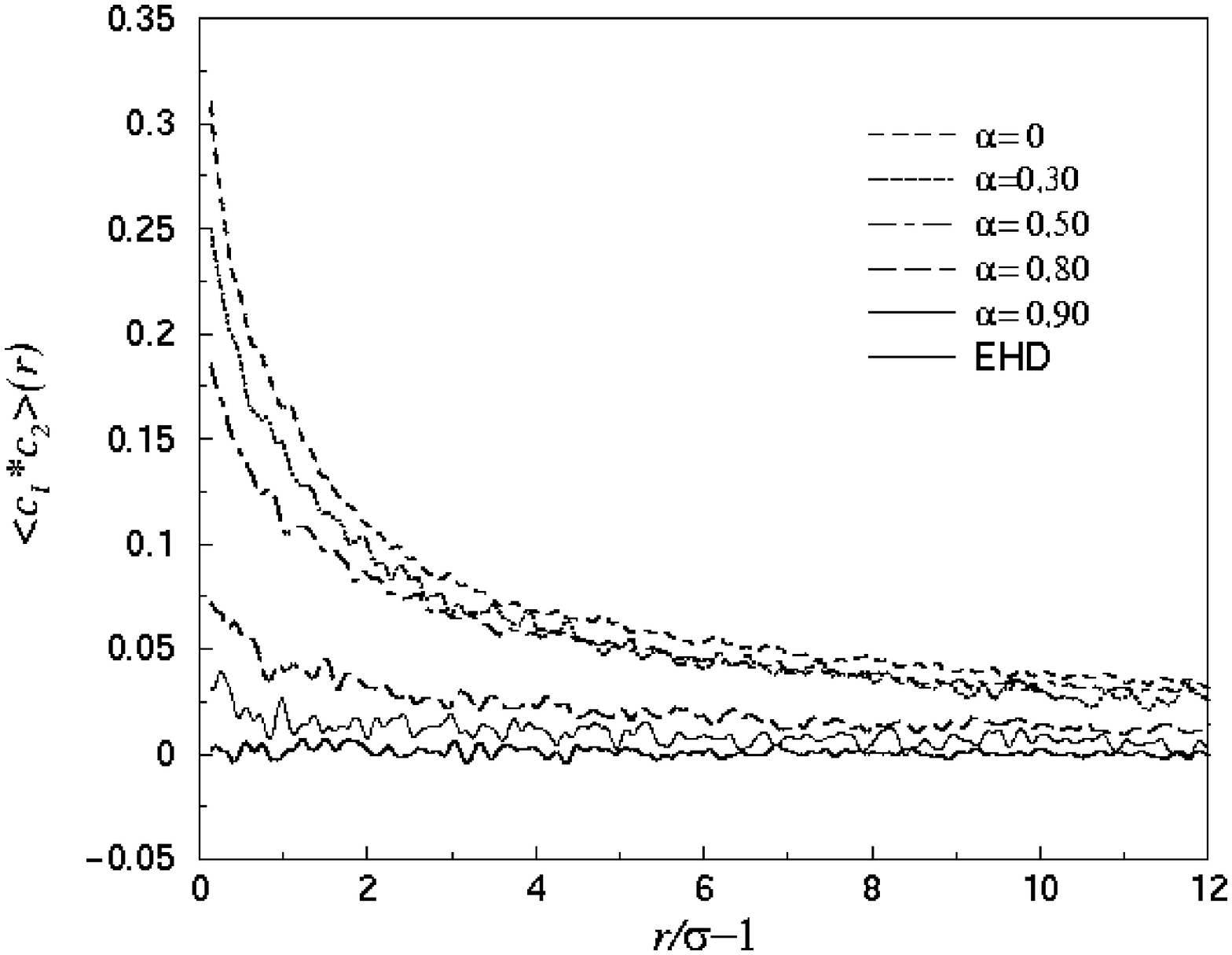,height=6.5cm,angle=0}
%\centerline{\hspace{8cm}\psfig{figure=./vpar_prof_10.eps,height=8cm}}
%\vspace{-8cm}
%\centerline{\hspace{9cm}\psfig{figure=./ln_vpar_11.eps,height=7.5cm}}
\caption{
\label{fig:v1v2r}}
\end{figure}
\end{center}

\begin{center}
\begin{figure}[h]
\vspace{-.5cm}
\epsfig{figure=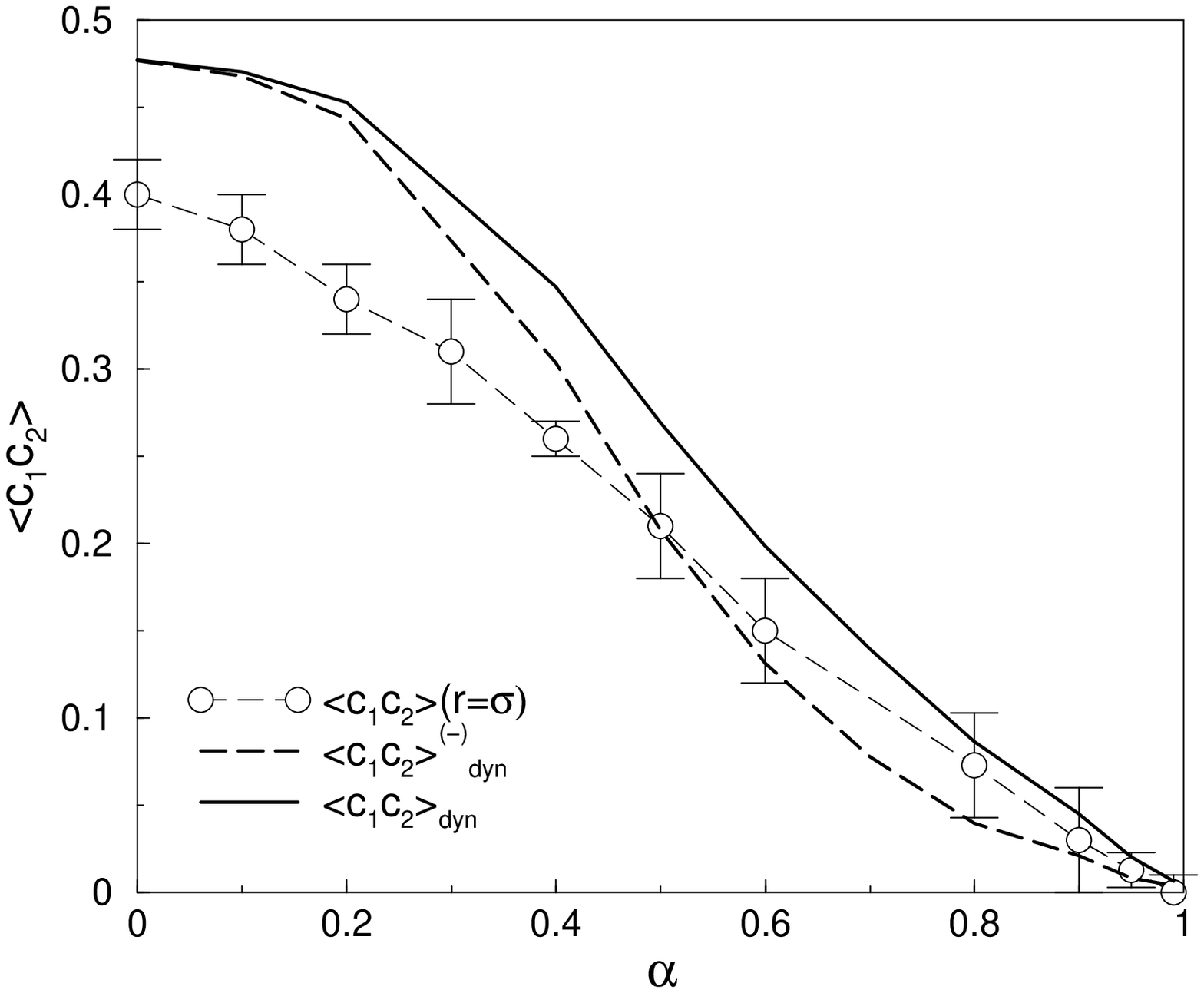,width=7.5cm,angle=0}
\caption{
\label{fig:v1v2sigma}}
\end{figure}
\end{center}

We show in Fig. \ref{fig:coll1} different velocity collisional
averages at $\phi=0.05$. In the simulations, these quantities are
obtained by averaging over successive collision events in the
steady state. We first observe that the simulation results in
Fig. \ref{fig:coll1}
%with a blow up in Fig. \ref{fig:coll2b}
approach for $\alpha\to 1$ the analytic results for elastic
spheres, calculated in (\ref{eq:averages}). At small
inelasticities, the simulation data follow the trends of the
theoretical prediction with systematic deviations depending on
the quantity considered. For instance, the behavior of the center
of mass motion $\langle G^2\rangle_{\hbox{\scriptsize coll}}  $
is close to the analytical prediction of     (\ref{eq:averages})
in the whole range of $\alpha$ values. This indicates that the
center of mass velocity ${\bf G}$ is not correlated with the
relative velocity ${\bf g}$.
 Consequently, $f^{(2)}({\bf c}_1,{\bf c}_2,\bsigma) $  in the
collisional average     (\ref{eq:collav}) factorizes, and we can
expect the contributions in numerator and denominator in
(\ref{eq:collav}) coming from ${\bf G}-$integrations to cancel.
Consistent with this behavior, we observe that the two curves in
Fig. \ref{fig:coll1} , $\langle c_1^2 \rangle_{\hbox{\scriptsize
coll}}$ (labeled by circles) and $\langle {\bf c}_1 \cdot {\bf
c}_2 \rangle_{\hbox{\scriptsize coll}}$ (labeled by squares), are
symmetric around $1/2$. In Eqs. (\ref{a16}) and
({\ref{a18}) of the appendix these quantities have been expressed
in reduced moments,
\ba \label{a24}
\langle {\bf c}_1 \cdot {\bf c}_2
\rangle_{\hbox{\scriptsize coll}} &=&
{\frac{1}{2}}- {\frac{3}{4}}\frac{ b_{31}}{b_{11}}
\nonumber \\
\langle
c_1^2 \rangle_{\hbox{\scriptsize coll}} &=& {\frac{1}{2}}+
{\frac{3}{4}}\frac{b_{31}}{b_{11}},
\ea
where
\be \label{a25}
b_{nm} = \frac{M_{nm}(T)}{M^{\rm E}_{nm}(T)} =
\left( \frac{T_{\rm E}}{T} \right)^{n/2} B_{nm}.
\ee
The reduced moments have been measured independently (see
Figs. \ref{fig:Bnn} and \ref{fig:bnm}), and used to calculate the expressions
(\ref{a24}) and (\ref{a25}). The results have been plotted in
Fig. \ref{fig:coll1} as dashed and dashed-dotted lines, which agree very well
with the direct measurements of these quantities as collisional
averages, shown in Fig. \ref{fig:coll1} respectively as squares and circles.
In deriving (\ref{a24}) and (\ref{a25}) we have again used that
the velocity variables ${\bf G}$ and ${\bf g}$ are statistically
uncorrelated. The present results strongly support this
assumption.

The correlation $\langle \hat{{\bf c}}_1 \cdot \hat{{\bf c}}_2
\rangle_{\hbox{\scriptsize coll}}= \langle \cos \psi_{12}
\rangle_{\hbox{\scriptsize coll}}$, also plotted in Fig. \ref{fig:coll1},
cannot be expressed in $B$-moments. However, the approximate
  relation already employed to show that
$\langle\cos \psi_{12}\rangle_{\hbox{\scriptsize coll}} \simeq
\langle {\bf c_1 \cdot c_2}\rangle_{\hbox{\scriptsize coll}}/
\langle c^2\rangle_{\hbox{\scriptsize coll}}$ in section
\ref{subsec:vel}, holds for the simulation
data over the whole range of inelasticities. As the system
becomes more inelastic, the typical ``temperature'' of colliding
particles (defined as the collisional average $\langle
c^2\rangle_{\hbox{\scriptsize coll}}$) decreases and even becomes
lower that the unconstrained average $\langle c^2 \rangle$ that
defines the temperature. On the other hand, as already noted below
(\ref{eq:averages}), $\langle c^2\rangle_{\hbox{\scriptsize
coll}} = 5/4>1$ in the elastic limit. This decrease of $\langle
c^2\rangle_{\hbox{\scriptsize coll}}$ is directly related to the
increase of the small $g$-portion of phase space where molecular
chaos is violated. At small $\alpha$, most of the collisions
occur between particles with small and even vanishing relative
velocities. An extreme example is the inelastic collapse,
mentioned in the introduction.

The correlation function $\langle {\bf c}_1 \cdot {\bf c}_2/g
\rangle_{\hbox{\scriptsize coll}}$ for the freely evolving IHD
fluid has been simulated by Soto and Mareschal, and was shown to
be small, but non-vanishing \cite{SM}.

We have also investigated $r-v-$correlations by measuring the
expectation value of $\langle {\bf c}_1\cdot {\bf c}_2\rangle(r)$
for two particles separated by a distance $r$, as defined in
(\ref{a21}). The results are shown in Fig. \ref{fig:v1v2r} and
Fig. \ref{fig:v1v2sigma}. The plot shows an intermediate range of
$r$ values with an exponentially decaying correlation. It is
again of interest to compare the extrapolation of the static
correlation $\langle {\bf c}_1 \cdot {\bf c}_2
\rangle (r \to \sigma)$ with its dynamic counter part,
$\langle {\bf c}_1 \cdot {\bf c}_2 \rangle_{\rm dyn}$, calculated
at collision. The results, derived in (\ref{a22}) and (\ref{a23})
of the appendix, read for hard disks,
\be \label{a26}
\langle {\bf c}_1 \cdot {\bf c}_2 \rangle_{\rm dyn}=
 {\frac{1}{2}} \left( 1- \frac{b_{20}}{b_{00}} \right)
+   \frac{1-\alpha}{4}\, \frac{b_{22}}{b_{00}}.
\ee
The first term on the RHS represents the precollision part,
\be \label{a27}
\langle {\bf c}_1 \cdot {\bf c}_2 \rangle^{(-)}_{\rm dyn}=
{\frac{1}{2}} \left( 1- \frac{b_{20}}{b_{00}} \right).
\ee
Figure \ref{fig:v1v2sigma} compares the extrapolation $\langle {\bf
c}_1 \cdot {\bf c}_2 \rangle (r \to \sigma)$ (circles) of the
static correlation with its dynamic analogs (\ref{a26}) and
(\ref{a27}). The numerical data for both correlations agree well
for $\alpha \gtrsim 0.8$, but for $\alpha \lesssim 0.5$ the
dynamic correlation (solid line) is substantially  larger than
the static one. This is consistent with the difference between
$\chi^{(-)}$ and $Y^{(-)}$ observed in Fig. \ref{fig:grcontact}}.
For comparison the dynamic precollision correlation
(dashed line) is also shown . It should be noted that the
divergence of $f^{(2)}$ at small $g$ and small $\cos
\theta$ implies in particular that $B_{00} \gg B_{20} > B_{22}$,
so that (\ref{a26}) predicts that the dynamic correlation at
contact $\langle {\bf c}_1\cdot {\bf c}_2\rangle_{\rm dyn}$
should increase at $\alpha
\to 0$ and saturate close to $1/2$. By the same arguments its
precollision part  in (\ref{a27}) approaches the same limit. This
can be observed in Fig. \ref{fig:v1v2sigma}.

The velocity correlation $\langle {\bf c_1 \cdot
c_2}\rangle_{\hbox{\scriptsize coll}}$ in (\ref{a24}) involve the
reduced moments $b_{31}$ and $b_{11}$. Consistent with the
scenario, developed in subsection III C, the divergence of
$f^{(2)}({\bf c}_1, {\bf c}_2,{\bf \sigma})$ near $g=0$ and $\cos
\theta=0$ is largely suppressed in these higher moments, which
remain finite for $\alpha \to 0$, where $ b_{11} \simeq 4 b_{31}$.
Consequently $\langle {\bf c_1 \cdot
c_2}\rangle_{\hbox{\scriptsize coll}}$ does not approach the
value $1/2$ as $\alpha \to 0$, but a value close to 0.3, as can
be deduced from Fig. \ref{fig:coll1}.

%%%%%%%%%%%%%%%%%%%%%%%%%%%%%%%%%%%%%%%%%%%%%%%%%%%%%%%%%%%%%%%%%%

\subsection{Grazing collisions}
\begin{center}
\begin{figure}[h]
\vspace{-0.5cm}
\epsfig{figure=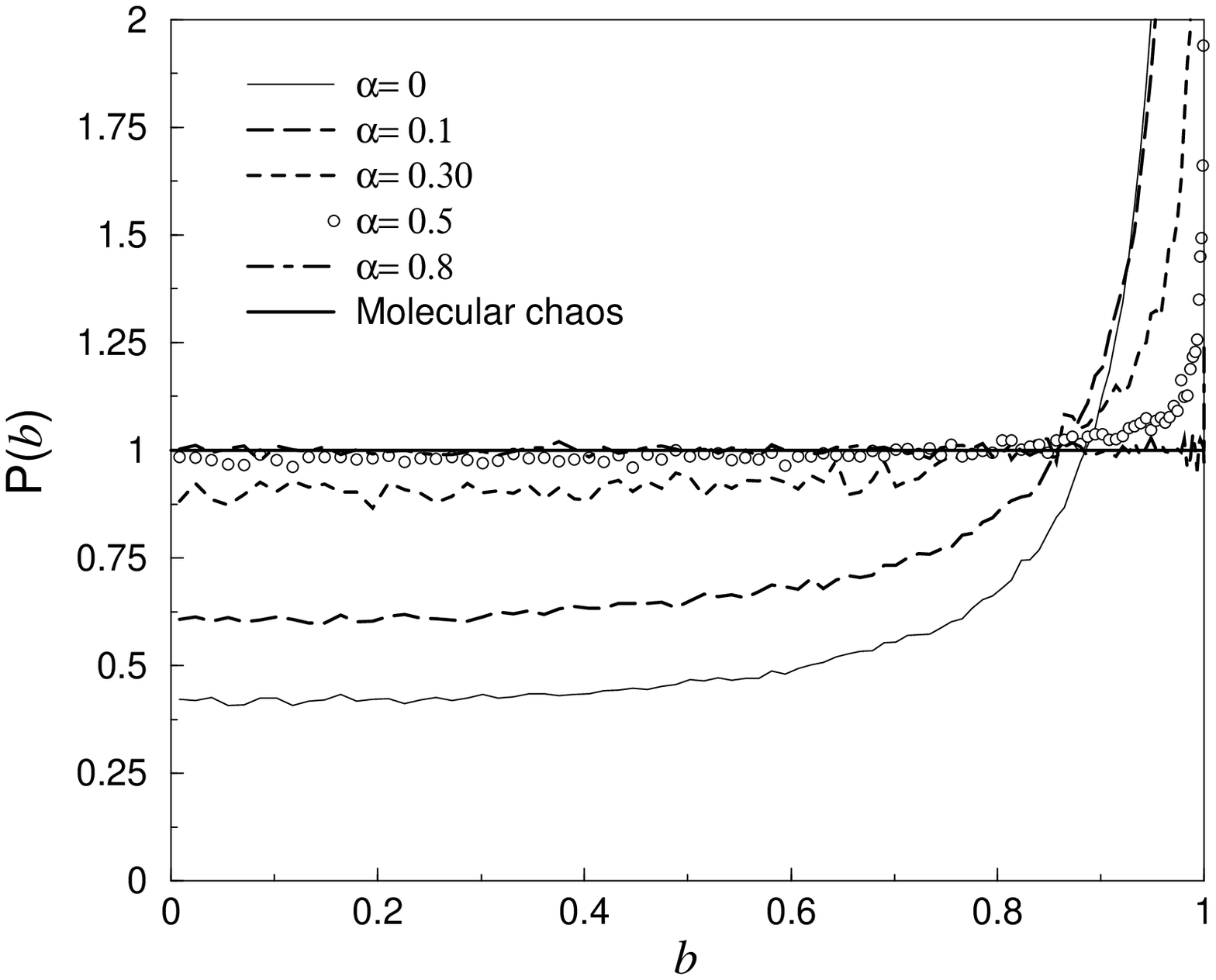,width=7.5cm,angle=0}
\caption{
\label{fig:b}}
\end{figure}
\end{center}

The data in Fig. \ref{fig:coll1} for $\langle {\bf c}_1\cdot{\bf
c}_2\rangle_{\hbox{\scriptsize coll}}  $, $\langle
\cos{\psi_{12}} \rangle_{\hbox{\scriptsize coll}},\langle
\sqrt{1-b^2}\rangle_{\hbox{\scriptsize coll}}$ and $\langle b\,
\rangle_{\hbox{\scriptsize coll}}$  clearly illustrate that
the violation of molecular chaos strongly increases with
increasing inelasticity. Consider first the average,
\be
\langle b \, \rangle_{\hbox{\scriptsize coll}}  =\int_0^1 {\rm d}b b P(b).
\ee
This average remains at a plateau value $1/2$ for $\alpha
\gtrsim 0.5$ , which is determined by the uniform distribution
$P(b)$ corresponding to molecular chaos in two dimensions. Recall
that the value $1/2$ holds regardless of the functional
form of the velocity distribution function $ f$. It is thus a good
probe for
 molecular chaos breakdown. Moreover, from its trend we can also
estimate the way in which such a breakdown takes place.
Specifically, as the inelasticity increases the average value
increases by about 50\%, which indicates a strong bias toward
grazing collisions. To illustrate this, we model the normalized
distribution of impact parameters as a uniform background and a
`half' delta-peak at $b=1$, i.e. $P(b)=1-p +2 p \delta (1-b)$,
where $p$ is the fraction of grazing collisions. This yields the
average $\langle b \,\rangle_{\hbox{\scriptsize coll}}
=\frac{1}{2}(1+p)$, which implies, according to Fig.
\ref{fig:coll1}, that at $\alpha=0.0, 0.1$ and 0.3 respectively a
fraction of 50\%, 35\% and 5\% is grazing at $\phi=0.05$. This qualitative
picture is supported in a more quantitative manner in Fig.
\ref{fig:b}, which shows the measured $P(b)$, which is strongly
peaked near grazing collisions $(b=1)$. At small inelasticity all
impact parameters are equally probable as expected on the basis
of molecular chaos, and consistent with Fig. \ref{fig:coll1}. Only
for $\alpha \lesssim 0.5$ deviations become significant: upon
decreasing the coefficient of restitution, collisions with a
larger impact parameter occur more frequently, implying an
increase of the frequency of grazing collisions. The behavior of
$P(b)$ is then fully consistent with the divergence of $f^{(2)}$
at small $\cos\theta$, discussed in subsection III C.

To avoid inelastic collapse for $\alpha \leq 0.5$ the
postcollision velocities of colliding pairs are rotated over a
small random angle as described in Refs.\ \cite{deltour,pre},
with the important restriction mentioned at the beginning of
section III. Alternative algorithms to avoid inelastic collapse
are described in Ref.
\cite{luding+mcnamara}. For $\alpha > 0.5$ no such rotation was applied. To
check if the deviations of the impact parameter for
$\alpha\lesssim 0.5$ are due to this applied rotation, we have
also performed simulations where even for $\alpha>0.5$ a random
rotation was applied. Regardless of the applied random rotation,
we found $\langle b \, \rangle_{\hbox{\scriptsize coll}} $ close
to $\frac{1}{2}$ for $\alpha\gtrsim 0.5$. Both Figs.
\ref{fig:coll1} and \ref{fig:b} show that for $\alpha \lesssim
0.5$ molecular chaos is strongly violated, and that the violation is weaker
in the small inelasticity regime.  The average $\langle
\sqrt{1-b^2} \rangle_{\hbox{\scriptsize coll}}$ supports the same
conclusions.

The data for $\langle {\bf c}_1\cdot{\bf c}_2
\rangle_{\hbox{\scriptsize coll}}  $ and
$\langle
\cos{\psi_{12}} \rangle_{\hbox{\scriptsize coll}}$ in Fig.\ref{fig:coll1}
are consistent with the predominance of grazing collisions at
large inelasticities. They show the average relative angle
between the velocities of the incoming particles, which has a
strong $\alpha-$dependence and no plateau value near the elastic
limit. Near $\alpha=1$ the particles are on average on approaching
trajectories with $\langle
\cos{\psi_{12}}\rangle_{\hbox{\scriptsize coll}}   \simeq -0.25$
and $\langle\psi_{12}\rangle_{\hbox{\scriptsize coll}}\simeq 105^o$. 
As $\alpha$ decreases, $\langle
\cos{\psi_{12}}\rangle_{\hbox{\scriptsize coll}}  $ increases
linearly to a value 0.50, while 
$\langle \psi_{12}\rangle_{\hbox{\scriptsize coll}} $ 
approaches $60^o$, at $\alpha=0$.  This corresponds to  collisions of more or
less parallel-moving pairs of particles, where faster particles
overtake slower ones.

\begin{center}
\begin{figure}[h]
\vspace{-0.5cm}
\epsfig{figure=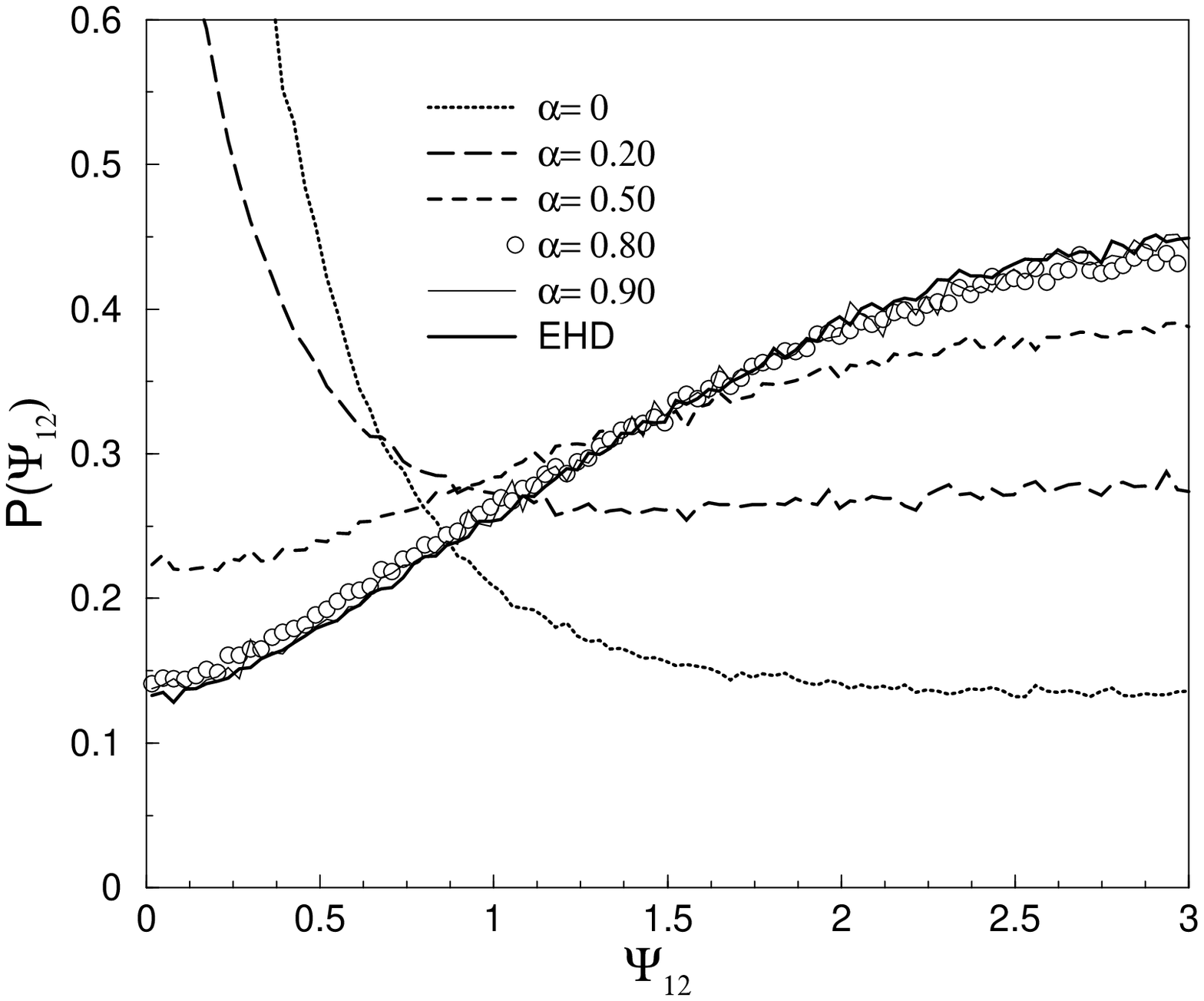,width=7.5cm,angle=0}
\caption{
\label{fig:theta}}
\end{figure}
\end{center}

\begin{center}
\begin{figure}[h]
\vspace{-0.5cm}
\epsfig{figure=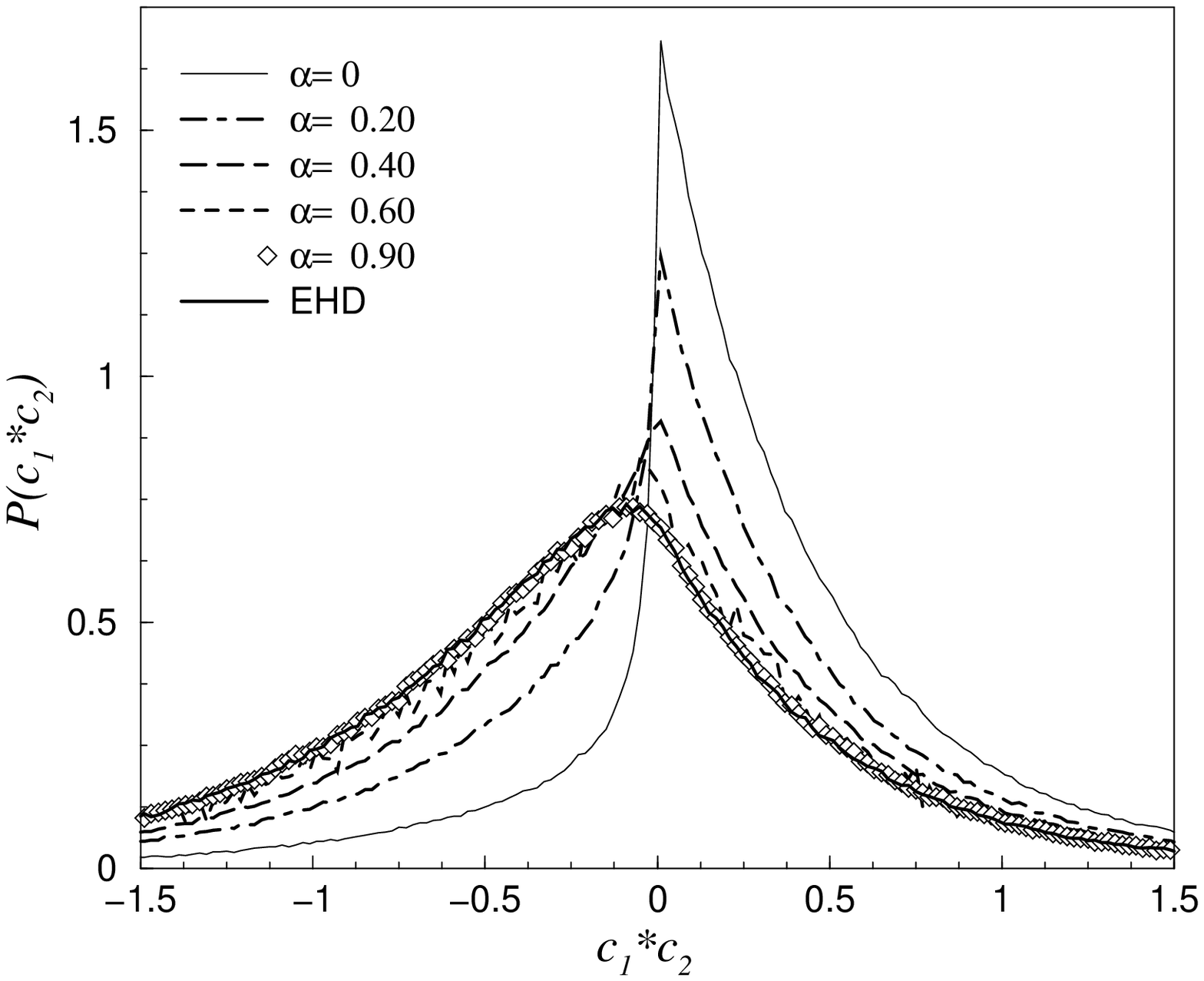,width=10cm,angle=0}
\caption{
\label{fig:Pc1c2}}
\end{figure}
\end{center}

Figs. \ref{fig:theta} and \ref{fig:Pc1c2} show the distribution of
relative orientations of incoming velocities. The distribution of
angles between the incoming particles ($\psi_{12}$) shows
moderate deviations from what is expected for an elastic system
in the range $0.5 \lesssim \alpha <1$. As an analytic expression
for elastic disks is not available, deviations are compared with
the simulation results for elastic hard disks (in the absence of
a random external force). At $\alpha=0.5$ the frequency of
collisions of parallel-moving particles is strongly increased, a
trend which is enhanced upon increasing the inelasticity.
Finally, the probability distribution $P({\bf c}_1\cdot {\bf
c}_2)$ is shown in Fig. \ref{fig:Pc1c2}. When the inelasticity
increases, this distribution becomes more peaked around the
origin, as the colliding particles on average move more slowly
relative to each other. In the mean time, the typical angle
$\psi_{12}$ decreases, which causes this peak to shift to
positive values.

%%%%%%%%%%%%%%%%%%%%%%%%%%%%%%%%%%%%%%%%%%%%%%%%%%%%%%%%%%%%%%%%%
%
\subsection{Particle- and noise-induced recollisions}
\label{sec:recoll}
The mechanism for the breakdown of molecular chaos in classical
fluids with conservative interactions are sequences of correlated
ring collisions, as discussed in the introduction.
The most simple ring collisions are the recollisions (1-2) (1-3)
(1-2) and cyclic collisions (1-2)(2-3)(3-1) or permutations thereof
\cite{berne}.

There is strong evidence that the effects of ring collisions are
considerably enhanced in fluids with dissipative interactions,
such as granular flows, where {\em relative} kinetic energy is
lost in binary collisions.
As a result the postcollision velocities
$\{\bv_1^\ast,\bv_2^\ast\}$ will be on average more parallel than
the precollision ones $\{\bv_1,\bv_2\}$\cite{mcnamara}, i.e.\ the trajectories
are less diverging than in the elastic case, and there is a much
larger $\{\br_3,\bv_3\}$ phase space, in which particle 3 will
knock, say, particle 1 back to recollide with particle 2.

This increase of phase space is confirmed by gathering
recollision statistics.
We have counted the fraction of recollisions as a function of
$\alpha$, as shown in Table \ref{table}.
The column labeled ${\cal R}_1$ (recollisions between two partners
mediated by a third particle) shows that at a packing fraction
$\phi=0.2$ in the elastic case ($\alpha=1$) only a fraction of
6.7\% of all collisions is a recollision.
This frequency gradually increases to about 15\% at $\alpha=0.4$.

In the randomly driven IHS fluid there is the additional effect
of noise induced recollisions that do not require the
intervention of other particles.
This type of recollision (denoted ${\cal R}_0$)
occurs with high probability when the
relative velocity after collision is so small that it can be
simply reversed by a random kick.
At $\alpha=0.6$ the frequency of noise-induced recollisions is
about 4\%, and it increases to 52\% at $\alpha=0$ (see column
${\cal R}_0$ in table \ref{table}).
 The effect is of importance at {\em all}
densities, because it does not require the mediation of a third
particle. Indeed, at a low packing fraction of 1\% and in the
completely inelastic case $\alpha=0$, the frequency of ${\cal
R}_0$-like events is still 34\%, while ${\cal R}_1$-like events
have dropped to 5\%. Moreover, we have verified that inclusion of
rotation-induced recollisions modifies most of the collisional
quantities we have analyzed, increasing their deviations with
respect to the molecular chaos prediction.

At present, more quantitative theories or estimates of the effect
of both types of recollisions and other ring collisions on the
short range behavior of the pair distribution function
$f^{(2)}(x_1,x_2)$ are lacking. A natural way to
incorporate the noise-induced recollisions into a
kinetic theory description  would be to include them into an
effective two-particle scattering operator, which transforms an
asymptotic precollision state of two independent particles into
an asymptotic postcollision state, without involving intermediate
two-particle scattering states, as in the present case. This may
lead to an instantaneous Boltzmann collision term (without memory
effects), provided the mean free time  and the time between
random kicks are very well separated (dilute gases). Such a
description would suppress the recollions of type ${\cal R}_0$,
and make the violation of molecular chaos less severe, say
comparable to the freely evolving IHD fluid.

%%%%%%%%%%%%%%%%%%%%%%%%%%%%%%%%%%%%%%%%%%%%%%%%%%%%%%%%%%%%%%%%%%%%%%%%%%%%%%%
%%%%%%%%
\subsection{Cold dense inhomogeneities}

\begin{center}
\begin{figure}[h]
\vspace{0.5cm}
\epsfig{figure=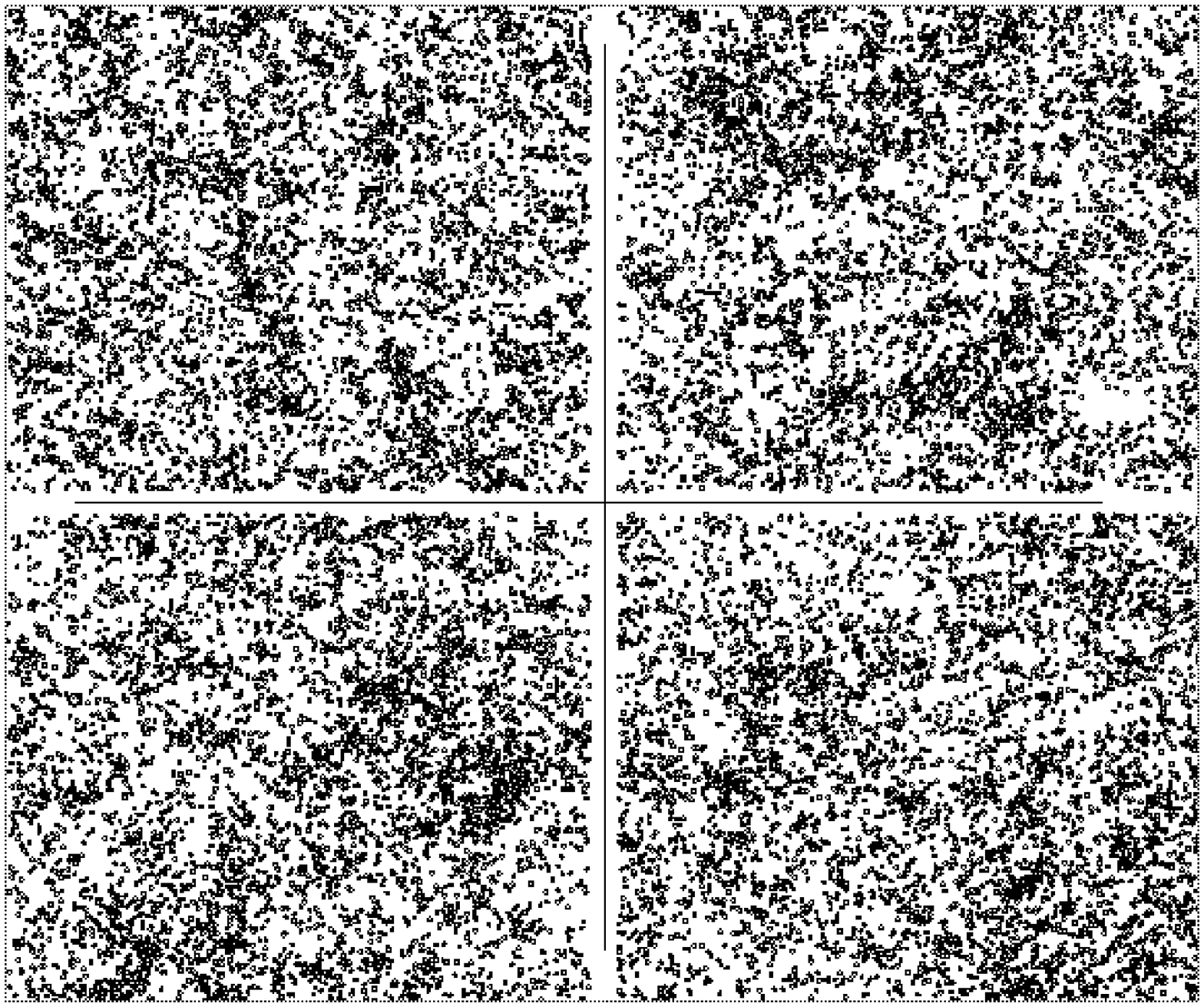,width=11cm,angle=0}
\caption{
\label{fig:clusterbis}}
\end{figure}
\end{center}

In Ref.\ \cite{pre} we have shown by analyzing the Fourier modes
of the granular hydrodynamic equations, which are valid for small
inelasticities (say $\alpha > 0.7$), that the NESS in a randomly
driven IHS fluid is linearly stable against spatial
inhomogeneities. Consequently, when observed over sufficiently
long times, the NESS should be spatially homogeneous. However, it was
also shown that the NESS exhibits strong fluctuations, resulting
in long range spatial correlations in density, flow field and
granular temperature. The observation of density inhomogeneities
for large inelasticities has already been reported by Peng and
Ohta \cite{peng}. These density inhomogeneities, as shown by the
snapshot of the density in Fig. \ref{fig:clusterbis}, are not
quasi-static, as in the freely evolving case
\cite{goldhirsch+zanetti,deltour,noijeprl,mcnamara}, but seem to
behave as dynamic assemblies of particles that dissolve and
re-assemble again. Also for a uniform shear flow,
dynamical density inhomogeneities
have been reported\cite{tan}.
The existence of density inhomogeneities was
already suggested by the static pair distribution functions
$g(r)$, which showed an enhancement of the first few maxima as
compared to their elastic values (see Fig.
\ref{fig:gr}).

 In Fig. \ref{fig:coll1} we show that the mean energy
$\langle c_1^2\rangle_{\hbox{\scriptsize coll}}  $, of particles
aiming to collide, is above the  mean, $\langle c^2 \rangle =1$,
for small inelasticity. It decreases from its elastic value $5
\langle c^2\rangle /4$ with decreasing
$\alpha$, then crosses the mean value value  $\langle c^2\rangle
=1$ at  $\alpha\simeq 0.2$, and further decreases to
approximately  $0.7
\langle c^2\rangle$ at $\alpha=0$.

It is interesting to observe that in the strong dissipation
range, the mean kinetic energy or granular temperature of
particles that are about to collide is {\em lower} than the
average temperature. We combine this observation with Figs. 3a,b
of Peng and Ohta \cite{peng}, which show that essentially all
collisions occur inside "cold" regions of high densities.
This last observation
applies even more so to  undriven IHS fluids
\cite{goldhirsch+zanetti,luding+herrmann}.
We expect that, also in the randomly driven IHS fluid, the
majority of collisions takes place inside cold high density
regions.

\begin{center}
\begin{figure}[h]
\centerline{\hspace{-8cm}\psfig{figure=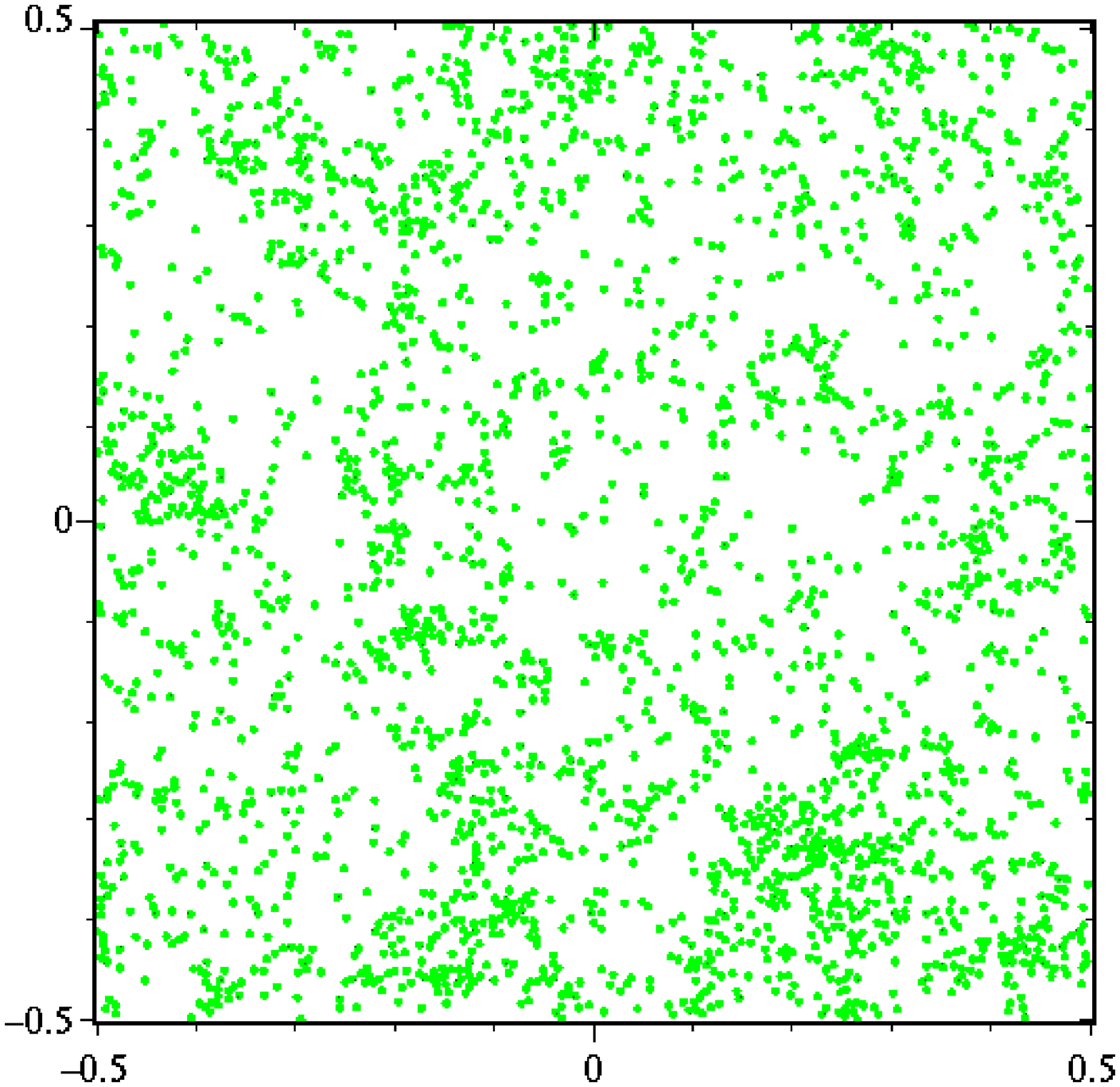,
height=7cm}}
\vspace{-7cm}
\centerline{\hspace{8cm}\psfig{figure=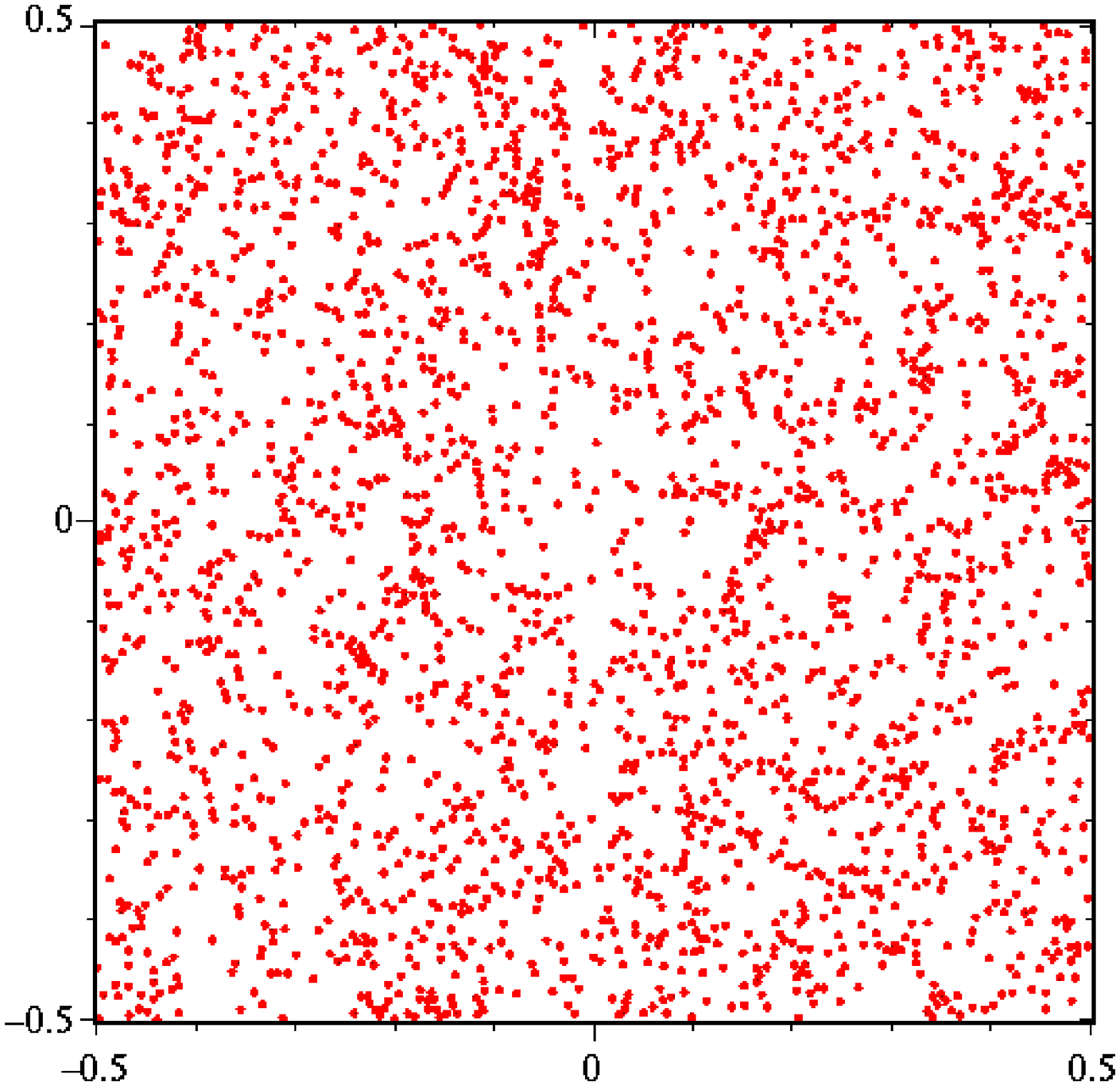,
height=7cm}}
\vspace{1.5cm}\caption{
\label{fig:densitysnapshot}}
\end{figure}
\end{center}

If the predominance of {\em cold} particles in strongly inelastic
collisions, $\langle c^2\rangle_{\hbox{\scriptsize coll}}
<\langle c^2\rangle$ is indeed a signal for the appearance of
density inhomogeneities, then Fig. \ref{fig:coll1} suggests that
at a packing fraction $\phi=0.05$ density inhomogeneities may
occur for $\alpha\lesssim 0.2$. This is indeed confirmed by
the snapshots in Figs. \ref{fig:clusterbis}. In Fig.
\ref{fig:densitysnapshot} we illustrate the existence of cold
inhomogeneous  dense regions for $\alpha=0.2$ and $\phi=0.2$. The
particles with a less (more) than median kinetic energy are shown
on the left (right). The formation of inhomogeneities is more
clear for the colder particles. The temporal evolution of these
regions show that they dissolve after some time, while new
inhomogeneous regions appear.
 The formation of ``living'' inhomogeneous regions can be
understood using the hydrodynamic picture put forward in
\cite{pre}, where it was shown that the structure factor behaves as
$S({\bf k}) \sim k^{-2}$, implying density correlations {\em
increasing} with distance as $\ln(r)$ in two dimensions. These
long range spatial correlations induce a slowing down of the
dynamics, as in critical  phenomena. This, in turn, implies the
slow decay of density perturbations, that could lead to visible
density inhomogeneities as the kicking frequency is reduced (in
this respect, see Refs.\cite{puglisi}). We can also expect that
upon decreasing the forcing frequency, the dynamics should be
closer to its ``free cooling'' counterpart so that well defined
clusters are then likely to appear.

\begin{center}
\begin{figure}[h]
%\centerline{\psfig{figure=./histovelscaled.eps}}
\centerline{\hspace{-8cm}\psfig{figure=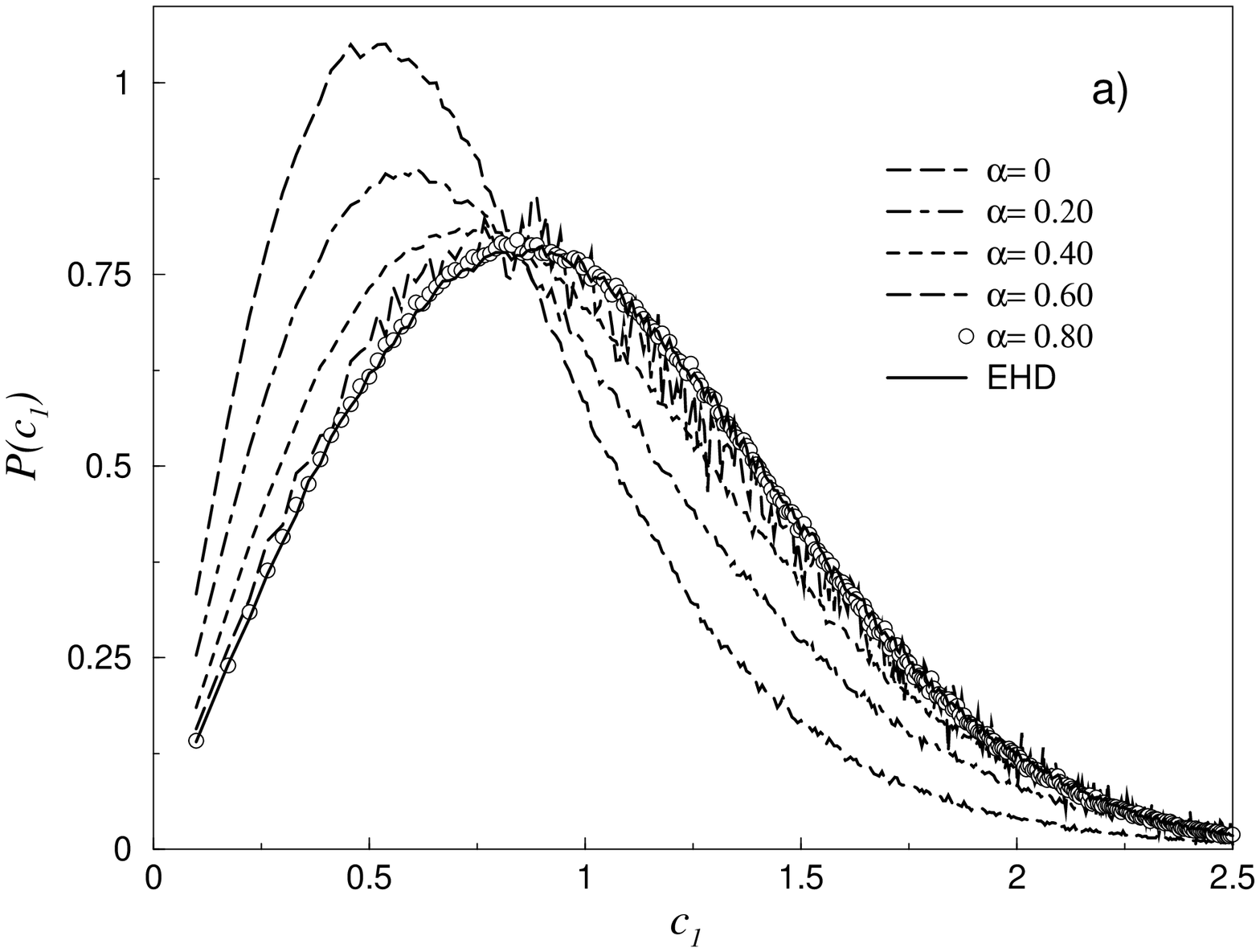,height=6cm}}
\vspace{-6.1cm}
\centerline{\hspace{8cm}\psfig{figure=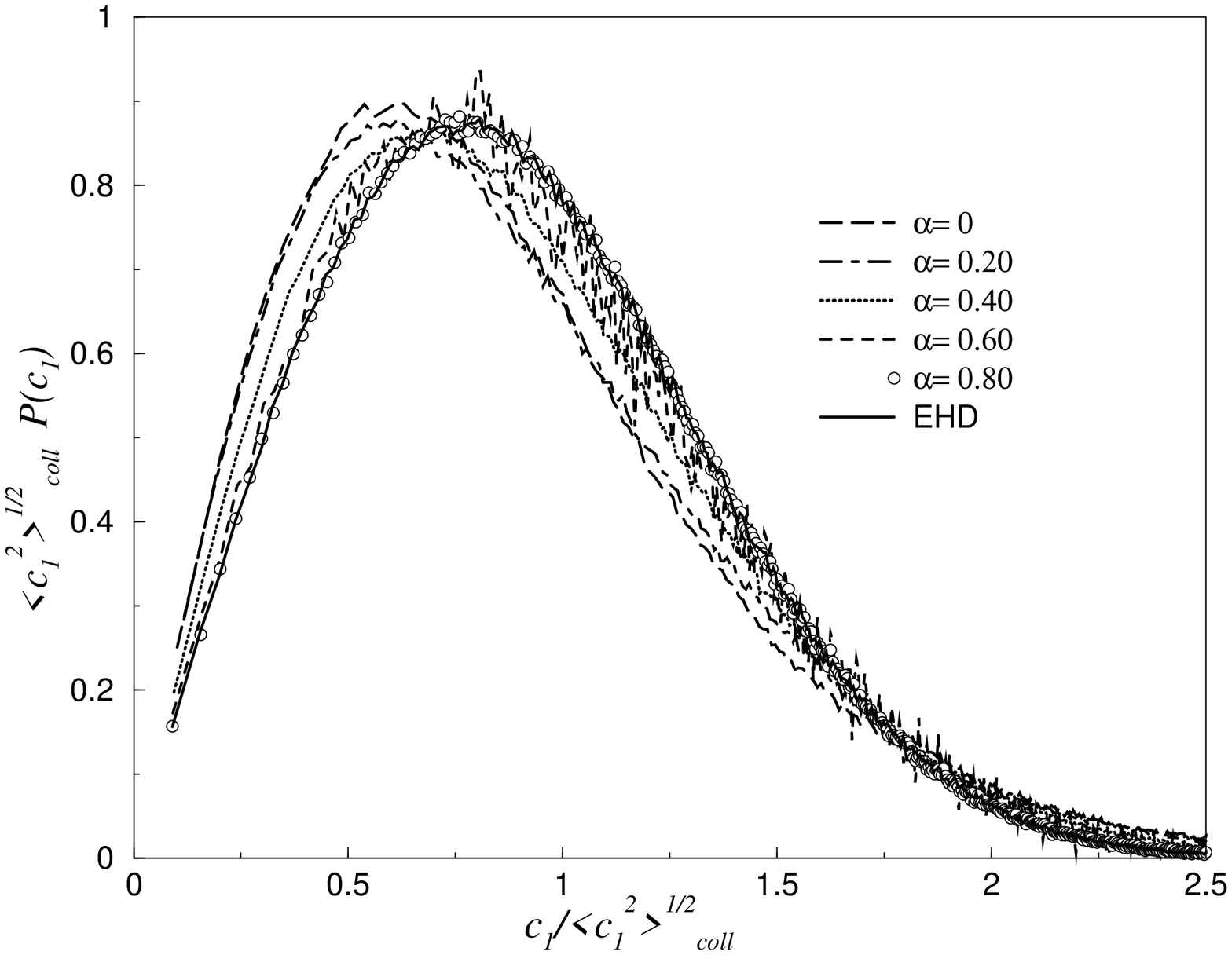,height=6cm}}
\caption{
\label{fig:Pc_1}}
\end{figure}
\end{center}

More details about the predominance of cold particles, among those
involved in collisions, can be seen in Fig. \ref{fig:Pc_1}, which
shows the constrained probability distribution $P(c)$, defined in
subsection II C and obtained from MD-simulations at different
inelasticities. For $\alpha \lesssim 0.5$ the distribution has
significantly shifted to smaller impact velocities. For the
completely inelastic case, collision events involving
``immobile'' particles are more than twice as frequent as for the
elastic case. The second moment of the distribution displayed in
Fig. \ref{fig:Pc_1} decreases when increasing the inelasticity. In
fact, all functional forms with simulation data at different
$\alpha$ can essentially be collapsed onto a single universal
curve (the elastic one) by plotting $\sqrt{T(\alpha)}
P(c|\alpha)$ as a function of $c/\sqrt{T(\alpha)}$, where
$T(\alpha)=\langle c^2\rangle_{\hbox{\scriptsize coll}}  $ is the
mean temperature of a particle at collision. The collapse plot is
shown in Fig. \ref{fig:Pc_1}b. This data collapse confirms the
concept of cold dense regions dominating the energy dissipation.
This could point to a possibly relevant two fluid picture  of a
``hot'' dilute background gas coexisting with continuously
rearranging configurations of "cold" dense regions.

%%%%%%%%%%%%%%%%%%%%%%%%%%%%%%%%%%%%%%%%%%%%%%%%%%%%%%%%%%%%%%%%%%%%%
\section{Conclusion}
\label{sec:conl}
We have performed extensive MD simulations to study the kinetic
properties
and short range correlations in the non-equilibrium steady state
of a randomly driven fluid of inelastic hard disks, as a model
for fluidized granular material.
The MD results have been compared with kinetic theory predictions
derived from the Enskog-Boltzmann equation, properly modified
with a Fokker-Planck diffusion term $\xi_0^2 (\partial/\partial
\bv)^2$ to account for the applied random driving force\cite{granmat}.

It appears that the kinetic theory predictions, based on
molecular chaos, are essentially in agreement with the MD results
for small inelasticities ($\alpha\gtrsim 0.5$) at $\phi=0.05$.
For larger inelasticities the deviations from the molecular chaos
predictions start to become manifest: the radial distribution
function at contact differs strongly from its local equilibrium
form; there is a predominance of grazing collisions. When
increasing $\phi$, the effects of the inelastic collisions become
relevant at smaller inelasticities; e.g. at $\phi=0.2$ 
and $\phi=0.5$, we observe already significant deviations for $\alpha\leq 0.7$.

To avoid inelastic collapse of the system at low
$\alpha$, we have implemented a modified rotation procedure (see
the beginning of section III). In its original version, this
procedure induces dramatic violations of molecular chaos. It
could then be argued that the important deviations of low order
reduced moments $B_{nm}$ are also spurious consequences of the
above rotation procedure. However, we checked that circumventing
the collapse by applying elastic collisions when the relative
velocity of a pair is below a certain
cutoff\cite{luding+mcnamara}, also induces very important
violations of molecular chaos (quantified by $B_{00}$ for
instance), unless the cutoff is chosen unphysically high.

Sequences of ring collision processes, which lead to the breakdown
of  molecular chaos in classical fluids with conservative
interactions, are strongly enhanced in fluids with dissipative
interactions, like rapid granular flows. We have analyzed how
molecular chaos is broken, i.e. essentially only through pairs of
colliding particles at very small relative velocities. This means
that molecular chaos is violated only in a small portion of phase
space, implying that only certain physical properties will be
sensitive to this violation. This explains why quantities like
the collision frequency, or the pair distribution function at
contact are very sensitive to the inelasticity parameters, while
others like the pressure or the energy dissipation rate are well
approximated by their Enskog prediction. Disentangling the
effects of hard disk and noise-induced correlations remains an
interesting point to explore. The studies performed in a freely
evolving IHS fluid also shows the predominance of grazing
collisions at long times. The fact that we have observed an
analogous behavior for this homogeneous steady state indicates
that the mechanism of breakdown of molecular chaos in granular
fluids through grazing collisions is generic for this type of
fluids.

The extra feature of noise-induced recollisions, which do not
require mediation of a third particle, will further enhance the
violation of the molecular chaos assumption. A natural way to
develop a kinetic theory for randomly driven fluids, thereby
presumably restoring the validity of the molecular chaos
assumption in the dilute gas case, could be to include the
noise-induced recollisions in an effective two-particle scattering
operator. It would be of interest to study its properties, either
analytically or by simulating a two-particle inelastic collision
in the presence of external noise. An additional theoretical
complication here is the validity of the Boltzmann equation
(\ref{eq:fp}) with Fokker-Planck diffusion term due to the fact
that there are two limits involved when dealing with hard spheres
in combination with external white noise. The actual properties
of the effective collision operator depend on the order in which
both limits are taken. In the simulations, one always takes the
hard sphere limit first, while the white noise is approximated by
discrete kicks which are applied to the particles at discrete
times.

In Ref.\ \cite{pre} we have calculated the equal-time spatial
correlations of the fluctuations in the hydrodynamic densities in
the NESS. Here we have focused on the dynamic properties of these
enhanced fluctuations, in particular of the dynamic
inhomogeneities  observed in the density field. The collisional
velocity moments, introduced in section
\ref{sec:predictions} and measured in MD simulations, reveal that
the dense regions consist mostly of particles colder than average.
This is clearly shown in the velocity distribution $P(v|\alpha)$
of particles which are about to collide.

The MD simulations have been carried out in the limit in which
the  time interval between the external random kicks is much
shorter than the {\it mean} free time between collisions. In this
limit, regions with density larger than average are not seen to
survive for a long time. Rather, they form, dissolve and reappear
elsewhere. The spatial correlations analyzed in
\cite{pre} show long-range correlations, which imply also a
slowdown in the temporal decay of density perturbations.
Therefore, we expect than the decrease of the kicking frequency
will be accompanied by the appearance of apparent clusters. This
fact, together with the shape modification of the  velocity
distribution  $P(v|\alpha)$ (see Fig. \ref{fig:Pc_1})
suggests the picture of a two fluid
model, in which a ``hotter'' more dilute background gas coexists
with continuously rearranging configurations of ``cold'' dense
clusters. This point remains open for subsequent investigation; for
example, it would be interesting to analyze separately the collisional
 statistics in the dense and dilute regions to assess the role of
density fluctuations.

%%%%%%%%%%%%%%%%%%%%%%%%%%%%%%%%%%%%%%%%%%%%%%%%%%%%%%%%%%%%%%%%%%%%%%%%%
\renewcommand{\theequation}{A.\arabic{equation}}
\setcounter{equation}{0}
\section*{Appendix: Reduced moments $B_{nm}(T)$}
In the body of the paper we have considered the collisional
averages $\langle  g^n |\cos\theta|^m\rangle_{\hbox{\scriptsize
coll}}  $ and the moments $M_{n,m} (T)$ and  $B_{n,m} (T)$. We
first list the Enskog values of these quantities, which have have
been calculated from its definitions, given below (\ref{Mnm}),\,
i.e.
\ba
&M\enu_{nm} (T_{\rm E}) = v_{\rm E}^n \chi_{\rm E} \, 2^{n/2}\,
\frac{\Gamma((d+n)/2)\Gamma((m+1)/2)}
{\sqrt{\pi}\Gamma((d+m)/2)}  & \label{a3}  \\
& \langle g^n |\cos \theta|^m
\rangle_{\hbox{\scriptsize coll}} \eno = 2^{n/2}
\frac{\Gamma((d+n+1)/2)\Gamma((m+2)/2)}{ \Gamma((d+m +1)/2)} &.
\label{MEnm}
\ea
Many physical quantities of interest can be expressed in terms of
reduced moments $B_{nm}$, as already illustrated in subsection II
C for $\chi^{(-)},\omega,p$ and $\Gamma$. Analogous relations
hold for the velocity moments $\langle
g^n\rangle_{\hbox{\scriptsize coll}}$, which are proportional to
$ M_{n+1,1}$. This yields
\be \label{a10}
B_{n+1,1} =\frac{\omega}{{\omega}_{{\rm E}}} \times
\frac{\langle v_{12}^n\rangle_{\hbox{\scriptsize coll}}}{\langle
v_{12}^n\rangle_{\hbox{\scriptsize coll}}\eno} =
\frac{\omega}{{\omega}_{{\rm E}}} \times
\frac{\langle g^n\rangle_{\hbox{\scriptsize coll}}}{\langle
g^n\rangle_{\hbox{\scriptsize coll}}\eno} \left(
\frac{T}{T_{{\rm E}}}\right)^{n/2} ,
\ee
where the denominator has been calculated in (\ref{MEnm}).

Velocity correlations between nearby particles can also be
expressed in the reduced moments $B_{nm}(T)$. First consider the
constrained averages $ \langle {\bf c}_1 \cdot {\bf
c}_2\rangle_{\hbox{\scriptsize coll}} $,  defined in
(\ref{eq:averages}). They contain $\langle G^2
\rangle_{\hbox{\scriptsize coll}} $, which equals d/4 from the MD
simulations, in agreement with (\ref{eq:averages}) (see
Fig. \ref{fig:coll1} of section III). The center of mass velocity
${\bf G}$ is consequently uncorrelated with the relative
velocity, and independent of the inelasticity. Substitution of
$\langle G^2 \rangle_{\hbox{\scriptsize coll}}=d/4$ in
(\ref{eq:averages}) yields,
\ba  \label{a16}
\langle {\bf c}_1 \cdot {\bf c}_2\rangle_{\hbox{\scriptsize coll}}
&=& \frac{d}{4}
- \frac{1}{4} \langle g^2\rangle_{\hbox{\scriptsize coll}}
=\frac{d}{4} -
\frac{d+1}{4}\left(\frac{T_{\rm E}}{T}\right) \frac{B_{31}}{B_{11}}    \\
\langle c^2_1\rangle_{\hbox{\scriptsize coll}}    &=& \frac{d}{4}
 + \frac{1}{4} \langle g^2\rangle_{\hbox{\scriptsize coll}}    =\frac{d}{4}
 +\frac{d+1}{4} \left(\frac{T_{\rm E}}{T}\right) \frac{B_{31}}{B_{11}}\  .
\label{a18}
\ea
Similarly we find,
\ba \label{a19}
&&\left\langle {\bf c}_1 \cdot {\bf c}_2/g
\right\rangle_{\hbox{\scriptsize coll}} \quad =\quad
\left\langle{G^2/g}\right\rangle_{\hbox{\scriptsize coll}}
- \textstyle{\frac{1}{4}} \langle g
\rangle_{\hbox{\scriptsize coll}} \nonumber \\
&& = \frac{d}{4\sqrt{2}}\frac{\Gamma(d/2)}{ \Gamma ((d+1)/2)}
\sqrt{\frac{T}{T_{{\rm E}}}}\times \frac{B_{01}}{B_{11}}
\left\{1- \left(\frac{T_{\rm E}}{T}\right)
\frac{B_{21}}{B_{01}}\right\}
\ea
and
\ba \label{a20}
&&\left\langle {\bf c}_1 \cdot {\bf c}_2 /|g \cos \theta|
\right\rangle_{\hbox{\scriptsize coll}}
 = \left\langle G^2/ |g \cos \theta| \right
 \rangle_{\hbox{\scriptsize coll}}
- \textstyle{\frac{1}{4}} \left\langle g/|\cos \theta|
\right\rangle_{\hbox{\scriptsize coll}} \nonumber \\
 && \quad = \frac{d}{4} \sqrt{\frac{\pi}{2}}\,
\sqrt{\frac{T}{T_{{\rm E}}}} \times \frac{B_{00}}{B_{11}}
\left\{1- \left(\frac{T_{\rm E}}{T} \right)
\frac{B_{21}}{B_{01}}\right\}.
\ea
Note that the last two averages are vanishing in the elastic
case.

In the body of the paper we have considered the extrapolation of
the static correlation $\langle {\bf c}_1 \cdot {\bf c}_2\rangle
(r \to \sigma)$. Here we calculate its dynamic analog $\langle
{\bf c}_1 \cdot {\bf c}_2\rangle_{{\rm dyn}}$, obtained by
interchanging limits and replacing  $ f^{(2)}({\bf c}_1,{\bf
c}_2, {r})$ under the integral sign in Eq.(\ref{Guu}) by its
value at contact, $ f^{(2)}({\bf c}_1,{\bf c}_2, {\bsigma})$. We
proceed in the same fashion as in
Eqs.(\ref{eq:g-unc})-(\ref{unc-min}), and split the numerator in
(\ref{a21}) in a pre- and postcollision part, as done in
subsection II D. One finds after a lengthy calculation,

\be \label{a22}
\langle {\bf c}_1 \cdot {\bf c}_2\rangle_{{\rm dyn}}=
{\frac{d}{4}} \left\{1- \left(\frac{T_{{\rm E}}}{T}\right)
\frac{B_{20}}{B_{00}} \right\}
+\frac{1-\alpha}{4} \left( \frac{T_{{\rm E}}}{T} \right)
\frac{B_{22}}{B_{00}}\,.
\ee
Here the first term on the RHS is its precollision part, i.e.
\ba \label{a23}
\langle {\bf c}_1 \cdot {\bf c}_2\rangle^{(-)}_{\rm dyn}&=&
\langle {\bf c}_1 \cdot {\bf c}_2 |g \cos \theta|^{-1}
\rangle_{\rm coll}\,\,/\,\,
\langle  |g \cos \theta|^{-1} \rangle_{\rm coll}
\nonumber \\
&=& {\frac{d}{4}} \left\{1-
\left( \frac{T_{\rm E}}{T} \right)
 \frac{B_{20}}{B_{00}} \right\}.
\ea
In section III these quantities are compared with MD
simulations.

\section*{Acknowledgments}
E.T. and I.P. thank D. Frenkel and B. Mulder for the
hospitality of
their groups at AMOLF. T.v.N. thanks `Laboratoire de
Physique Th{\'e}orique'
of the `Universit{\'e} Paris-Sud, Orsay' where part of
this work was done.
We acknowledge stimulating discussions with H.J. Herrmann,
S. Luding, R. Soto, M. Mareschal and J. Piasecki.
T.v.N. and I.P. acknowledge support of the
foundation `Fundamenteel Onderzoek der Materie (FOM)', which is
financially supported by the Dutch National Science Foundation
(NWO).

\begin{table}
%\begin{center}
\begin{tabular}{|l|l|l|}
$\alpha$       &   ${\cal R}_0$ & ${\cal R}_1$ \\ \hline
     0            &     52    \%         &              18   \%\\
     0.4          &     14    \%         &              15   \%\\
     0.6          &      4    \%         &              15   \%\\
     0.95         &      0.15 \%         &               7   \%\\
     1.0          &      0    \%         &               6.7 \% \\
\end{tabular}
%\end{center}
\caption{Frequency of recollision events as a function of the
inelasticity (see text for the
definition of ${\cal R}_0$ and ${\cal R}_1$).
The packing fraction is
$\phi = 0.2$ and the system contains $N=5000$ disks.}
\label{table}
\end{table}

\begin{center}
{FIGURE CAPTIONS}
\end{center}

\begin{enumerate}

\item Kinetic temperature $T/T_{\rm E}$ and  collision frequency
$\omega/\omega_{\rm E}$ where $T_{\rm E}$ and $\omega_{\rm
E}$ are defined in Eqs. (\ref{eq:T-Ensk}) and (\ref{omega-Ensk}), for a
packing fractions $\phi=0.05$ and $\phi=0.2$.

\item Fourth cumulant as a function of the coefficient of
restitution. Comparison is made between the two-dimensional
version of  (\ref{eq:a2}) and MD results (circles) obtained for a
system of 10201 inelastic disks, measured at several densities
(see section III A).

\item Pair distribution functions $g(r)$  versus distance between the
particles at a packing fraction $\phi=0.2$. The arrow indicates
the value at contact for an elastic hard disc (EHD) fluid (from
Verlet and Levesque \cite{verlet}).

\item a) Static or unconstrained pair distribution functions at
contact $Y,Y^{(-)},Y^{(+)}$ , as extrapolated from the
corresponding pair distribution function $g(r \to
\sigma)$ at $\phi=0.05$, compared with the dynamic correlation $\chi^{(-)}$
at contact. The straight line corresponds to the EHD prediction
($\chi_{\rm E}= 1.084$). b) Ratio of dynamic to static correlation
$\chi^{(-)}/Y^{(-)}$, to be compared with 1, and the static ratio
$Y^{(-)}/ Y^{(+)}$, to be compared with the dynamic ratio
$\chi^{(-)}/\chi^{(+)} =\alpha$.

\item Pressure versus coefficient of restitution at a packing fraction
$\phi=0.05$. The simulations results (direct or through $B_{22}$)
are compared with molecular chaos prediction (\ref{eq:pressmc}),
where $\chi$ is either the static $Y^{(-)}$ in (\ref{sumY}), or
the dynamic $\chi^{(-)}$ in (\ref{eq:g-min}), or the Enskog
approximation  $\chi_{\rm E}$ in (\ref{eq:vlcs}), corresponding
to $B_{22} =1$.

\item Reduced moments $B_{nn}$ for $n=1,2,3$ as a function of
the restitution coefficient at $\phi=0.2$ and $N=5041$. A similar behaviour
is observed at lower densities.

\item Reduced moments $B_{0m}$ and $B_{n0}$ as a
function of $\alpha$ at $\phi = 0.2$ and $N=5041$.

\item Values of different collisional averages
obtained in MD simulations, as a function of $\alpha$ at $\phi
=0.05$ (see section II B for definitions). For $\alpha<0.5$ the
random rotation introduced to avoid inelastic collapse has a
maximum deviation angle of $2.5^o$. The symbols $b_{nm}$ are
defined in Eq. (\ref{a25}).

\item  Distribution of $\langle {\bf c}_1\cdot{\bf c}_2\rangle(r)$
as a function of the distance between the particles at
$\phi=0.05$.

\item Mean velocity-velocity correlation function at contact,
as extrapolated from $\langle {\bf c}_1\cdot{\bf c}_2\rangle(r)$
(previous figure), compared with the dynamic analogs $\langle
{\bf c}_1\cdot{\bf c}_2\rangle_{\rm dyn}$ and  $\langle {\bf
c}_1\cdot{\bf c}_2\rangle^{(-)}_{\rm dyn}$, defined in (\ref{a22})
and (\ref{a23}).

\item Distribution of the impact parameter $b$ for different
$\alpha$-values for $\phi = 0.05$.

\item Distribution of the relative orientation of the velocities
at collision ($\cos\psi_{12} \equiv \widehat{\bf
c}_1\cdot\widehat{\bf c}_2$) at a packing fraction of 5\%.

\item Distribution of relative velocities, ${\bf c}_1\cdot{\bf c}_2$, of
 colliding inelastic disks, at $\phi =0.05$.

\item
To illustrate the slow reorganization of density inhomogeneities, four
consecutive snapshots of the system are shown at $\alpha=0.1$,
$\phi=0.2$ and $N=5000$ (the full simulation box is displayed).
The time interval between two consecutive snapshots corresponds
to 50 collisions per particle.

\item Snapshot of a typical instantaneous configuration
of the system at $\alpha=0.2$, $\phi=0.2$ and $N=5000$. To
illustrate the existence of cold dense inhomogeneities, on the
left (right) the particles with a less (more) than median kinetic
energy \smash{${\cal E}^*$} are shown at real scale (i.e. the
cutoff \smash{${\cal E}^*$} is chosen such that there are exactly
half of the particles on each graph). Lengths on the $x$ and $y$ axes are
expressed in units of the simulation box length.

\item Velocity distribution of the colliding
particles at $\phi=0.05$ and $N=5041$: a) Original distribution;
b) scaled velocity distributions as a function of the rescaled
velocity $c/\sqrt{T(\alpha)}$ for different values of $\alpha$.

\end{enumerate}

\end{document}